\documentclass[aps,prl,reprint,groupedaddress,nofootinbib,longbibliography]{revtex4-1}
\usepackage{amsmath}
\usepackage{amsthm}
\usepackage{amssymb}
\usepackage{graphicx}
\usepackage{mathrsfs}
\usepackage{mathtools}
\usepackage{enumerate}
\usepackage{enumitem}
\usepackage[T1]{fontenc}
\usepackage[sc]{mathpazo} 
\usepackage{physics}
\usepackage{subfigure}
\usepackage{overpic}
\usepackage{epstopdf}
\usepackage{bm}
\usepackage{color,soul}
\usepackage{dcolumn}
\usepackage[bookmarks=false]{hyperref}
\usepackage{xcolor}
\usepackage{booktabs} 
\usepackage{makecell}
\usepackage{color,soul}
\usepackage[linesnumbered]{algorithm2e}
\hypersetup{
    colorlinks=true,
    citecolor=blue,
    linkcolor=blue,
    urlcolor=blue,
    pdfstartview=FitH,
    bookmarksopen=true
    }
\newtheorem{theorem}{Theorem}
\newtheorem{lemma}{Lemma}

\makeatletter 
    
\renewcommand\onecolumngrid{% <<<<<<
\do@columngrid{one}{\@ne}%
\def\set@footnotewidth{\onecolumngrid}% <<<<<<<<<<<<<<<<
\def\footnoterule{\kern-6pt\hrule width 1.5in\kern6pt}%
}

\renewcommand\twocolumngrid{% <<<<<<
        \def\footnoterule{% restore rule
        \dimen@\skip\footins\divide\dimen@\thr@@
        \kern-\dimen@\hrule width.5in\kern\dimen@}
        \do@columngrid{mlt}{\tw@}
}%

\makeatother    
\usepackage{cleveref}
\crefname{equation}{Eq.}{Eqs.}
\crefname{theorem}{Theorem}{Theorems}
\crefname{appendix}{Appendix}{Appendixes}
\crefname{figure}{Fig.}{Figs.}
\crefname{section}{Sec.}{Secs.}
\crefname{algorithm}{Algorithm}{Algorithms}

\begin{document}

%\title{Strict hierarchy of operationally realizable strategies in non-asymptotic quantum metrology}
%\title{Unveilling strict hierarchy of operationally realizable strategies by linking global estimation theory with local one}
%\title{Linking global estimations with local ones to reveal strict hierarchy of operationally realizable strategies}
\title{Strict hierarchy of optimal strategies for global estimations: Linking global estimations with local ones}
%\title{Mapping global parameter estimations into local ones via virtual imaginary time evolutions}

\author{Zhao-Yi Zhou}
\affiliation{Department of Physics, Shandong University, Jinan 250100, China}
\author{Jing-Tao Qiu}
\affiliation{Department of Physics, Shandong University, Jinan 250100, China}
\author{Da-Jian Zhang}
\email{zdj@sdu.edu.cn}
\affiliation{Department of Physics, Shandong University, Jinan 250100, China}

\date{\today}

\begin{abstract} 
A crucial yet challenging issue in quantum metrology is to ascertain the ultimate precision achievable in estimation strategies. While there are two paradigms of estimations, local and global, current research is largely confined to local estimations, which are useful once the parameter of interest is approximately known. In this Letter we target a paradigm shift towards global estimations, which can operate reliably even with a few measurement data and no substantial prior knowledge about the parameter. The key innovation here is to develop a technique, dubbed virtual imaginary time evolution, which establishes an equality between the information gained in a global estimation and the quantum Fisher information for a virtual local estimation. This offers an intriguing pathway to surmount challenges in the realm of global estimations by leveraging powerful tools tailored for local estimations. We explore our technique to reveal a strict hierarchy of achievable precision for different global estimation strategies and uncover unexpected results contrary to conventional wisdom in local estimations.
\end{abstract}

\maketitle

\textit{Introduction}.---Quantum metrology lies at the heart of quantum science and technologies, aiming to design optimal strategies for precisely estimating unknown parameters with limited resources. The burgeoning capabilities of quantum sensors have opened up possibilities of harnessing quantum mechanical effects like entanglement to yield quantum metrological advantages \cite{Dorfman2016RMP,Degen2017RMP,Pezze2018RMP}. This allows quantum metrology to push precision limits beyond the reach of classical methods, holding compelling promise for applications such as quantum imaging, quantum interferometry, and quantum thermometry \cite{Giovannetti2011NP,DemkowiczDobrzanski2015PO,Pirandola2018NP}.

The prototypical setting of quantum metrology is to estimate an unknown parameter $\theta$ carried by a quantum channel $\mathcal{E}_\theta$ with $N$ queries to $\mathcal{E}_\theta$. Various types of estimation strategies can be employed for this purpose: ($i$) parallel strategies \cite{Giovannetti2006PRL}, where these $N$ channels are applied simultaneously on a multipartite entangled state; ($ii$) sequential strategies \cite{Giovannetti2006PRL}, involving successive queries of the channels, possibly interspersed with unitary control operations; ($iii$) causal superposition strategies \cite{zhaoQuantumMetrologyIndefinite2020}, where the channels are probed in a superposition of different causal orders; ($iv$) general indefinite-causal-order strategies \cite{liuOptimalStrategiesQuantum2023}, encompassing the most general causal relations among the channels and including causal superposition strategies as special cases. 
A crucial yet challenging issue in quantum metrology is to ascertain the ultimate precision achievable in estimation strategies, which has recently motivated a vibrant activity \cite{Giovannetti2006PRL,PARIS2009IJQI,zhaoQuantumMetrologyIndefinite2020,Demkowicz_Dobrza_ski_2012,DemkowiczDobrzanski2014PRL,yangMemoryEffectsQuantum2019,Cheng2019PRA,altherrQuantumMetrologyNonMarkovian2021,Liu2022AQT,Zhang2022nQI,Ullah2023PRR,Kurdzialek2023PRL,liuOptimalStrategiesQuantum2023}. 

While there are two paradigms of estimations in quantum metrology \cite{PARIS2009IJQI},  local and global, the current research on the issue is largely confined to local estimations \cite{Giovannetti2006PRL,PARIS2009IJQI,zhaoQuantumMetrologyIndefinite2020,Demkowicz_Dobrza_ski_2012,DemkowiczDobrzanski2014PRL,yangMemoryEffectsQuantum2019,Cheng2019PRA,altherrQuantumMetrologyNonMarkovian2021,Liu2022AQT,Zhang2022nQI,Ullah2023PRR,Kurdzialek2023PRL,liuOptimalStrategiesQuantum2023}. This bias is partially attributed to the fact that the performance of local estimation strategies is characterized by the quantum Fisher information (QFI) \cite{braunsteinStatisticalDistanceGeometry1994,Zhang2020PRR} and many powerful tools are available for computing the QFI. However, unless dealing with a special class of probability models \cite{DemkowiczDobrzanski2015PO}, local estimation strategies require the parameter to be approximately known.
This severely restricts their applicability, excluding diverse situations where little is known about the parameter \textit{a priori} \cite{Gorecki2020PRL,Gorecki2022PRL}. Alternatively, the knowledge about the parameter may be acquired \textit{a posteriori} using a sufficiently large number of measurement samples \cite{BarndorffNielsen2000JoPAMaG,Gill2000PRA}, which, nevertheless, demands too much experimental effort.

Here we target a paradigm shift towards global estimations. Unlike local ones, global estimation strategies can operate reliably even with a few measurement data and no substantial prior knowledge about the parameter  \cite{Valeri2020nQI,Montenegro2021PRL,Rubio2021PRL,Mok2021CP}. This general applicability is highly valuable in realistic settings, given the limited capabilities of near-term quantum sensors \cite{Preskill2018Q,Jiao2023AQT}. Unfortunately, useful tools for evaluating the performance of global estimation strategies are currently lacking \cite{Meyer2023,Bavaresco2023}, raising technical obstacles in tackling the mentioned issue. Consequently, whereas significant advancements have been made towards fully understanding quantum metrological advantages in local estimations \cite{Giovannetti2006PRL,PARIS2009IJQI,zhaoQuantumMetrologyIndefinite2020,Demkowicz_Dobrza_ski_2012,DemkowiczDobrzanski2014PRL,yangMemoryEffectsQuantum2019,Cheng2019PRA,altherrQuantumMetrologyNonMarkovian2021,Liu2022AQT,Zhang2022nQI,Ullah2023PRR,Kurdzialek2023PRL,liuOptimalStrategiesQuantum2023}, the progress in global estimations has remained quite limited so far. 

The key innovation of this Letter is to develop a technique, dubbed virtual imaginary time evolution (ITE), which allows us to establish an equality between the information about the parameter gained in a global estimation and the QFI associated with a virtual local estimation.
We can therefore figure out the ultimate precision achievable in global estimation strategies by computing the QFI. Aided by our technique, we tackle the hierarchy problem on ultimate precision achievable in global estimation strategies, which stands out due to its vital role in understanding quantum metrological advantages but remains open until now \cite{Bavaresco2021,liuOptimalStrategiesQuantum2023}. Meanwhile, we uncover some unexpected results contrary to the conventional wisdom established for local estimations. 
The result of this Letter offers an intriguing pathway to surmount challenges in the realm of global estimations by leveraging powerful tools tailored in local estimations.

\begin{figure}
	\centering
	\includegraphics[width=\linewidth]{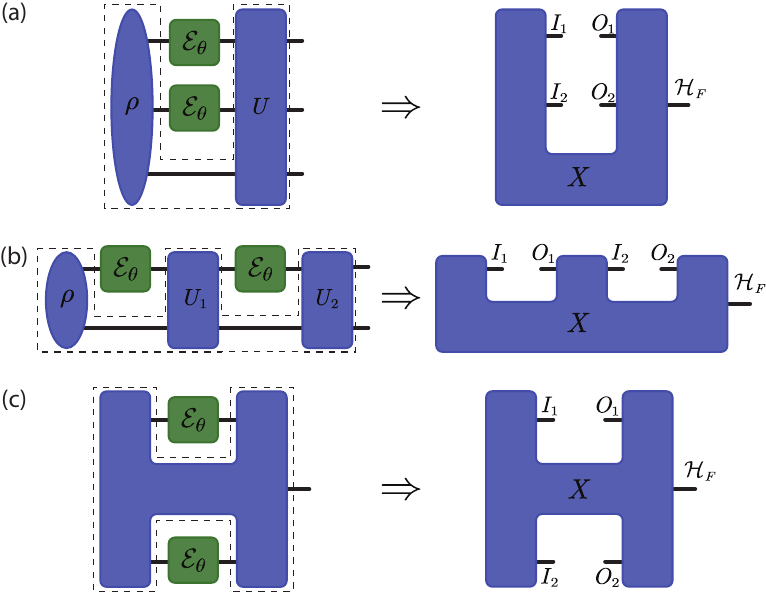}
	\caption{Schematic of (a) parallel strategies, (b) sequential strategies, and (c) general indefinite-causal-order strategies in the $N=2$ case. The left column illustrates that a strategy is an arrangement of physical operations such as initial state preparations and adaptive controls. This arrangement, when concatenated with the $N$ channels, produces an output state carrying information about $\theta$. The right column depicts that a strategy amounts to a supermap which, akin to completely positive maps, can be described by a positive semidefinite operator $X$. 
	}
	\label{fig:schematic_figure}
\end{figure}

\textit{Preliminaries}.---Let $\rho_\theta$ denote the state produced in a strategy. To estimate $\theta$, one needs to perform a positive operator-valued measure (POVM) $\{\Pi_x\}$ on $\rho_\theta$ and then post-processing the measurement outcome via an estimator $\hat{\theta}(x)$, where $x$ labels the outcome. The objective of a local estimation is to choose suitable $\{\Pi_x\}$ and (unbiased) $\hat{\theta}(x)$ to minimize
{the local variance}
$\mathrm{Var}\left[\hat{\theta}|\theta \right] =\sum_x{p\left( x|\theta \right) \left[ \hat{\theta}\left( x \right) -\theta \right] ^2}$,
where $p(x|\theta)=\mathrm{tr}\left(\Pi _x\rho_\theta\right)$. The quantum Cram\'{e}r-Rao bound reads 
$\mathrm{Var}\left[\hat{\theta}|\theta \right]\geq {1}\big/{\mathcal{I}\left[\rho_\theta \right]}$ \cite{braunsteinStatisticalDistanceGeometry1994}, where
\begin{eqnarray}\label{QFI1}
\mathcal{I}[\rho_\theta]=\mathrm{tr}\left(\rho_\theta L_\theta^2 \right)
\end{eqnarray}
denotes the QFI, with $L_\theta$ being the symmetric logarithmic derivative defined as the Hermitian operator satisfying 
\begin{equation}\label{QFI2}
\frac{d}{d\theta}\rho_\theta=\left(\rho_\theta L_\theta+L_\theta \rho_\theta\right)/2.
\end{equation}
Notably, the optimal measurement saturating the bound typically depends on $\theta$, implying that local estimations are useful only when $\theta$ is approximately known. Unlike local estimations, a global estimation aims to minimize {the global variance}  
$\mathrm{Var}[\hat{\theta}]=\int d\theta\sum_xp(x,\theta)[\hat\theta(x)-\theta]^2$ \cite{Note1},
where $p(x,\theta)=p(\theta)p(x|\theta)$ is the joint probability distribution of $x$ and $\theta$, with $p(\theta)$ denoting the prior probability distribution of $\theta$. It has been shown \cite{Personick1971ITIT,Helstrom1976,Macieszczak2014NJP,Rubio2019NJP,DemkowiczDobrzanski2020JPAMT,Sidhu2020AQS,Rubio2024} that $\mathrm{Var}[\hat{\theta}]$ is bounded by
\begin{equation} \label{eq:global_inequality}
\mathrm{Var}[\hat\theta] \ge \int d\theta p(\theta)\theta^2 - \mathrm{tr}(\bar\rho S^2).
\end{equation}
Here $\bar\rho=\int d\theta p(\theta)\rho_\theta$ is the averaged state and $S$ is a Hermitian operator satisfying 
\begin{equation}\label{S}
    \overline{\theta \rho }=\left( \bar{\rho}S+S\bar{\rho} \right)/2,
\end{equation}
with $\overline{\theta\rho}\coloneqq\int d\theta p(\theta)\theta\rho_\theta$. The inequality (\ref{eq:global_inequality}) can be saturated by choosing $\{\Pi_x\}$ as the projective measurement of $S$ and $\hat \theta(x)$ as the eigenvalues of $S$ \cite{Personick1971ITIT,Helstrom1976,Macieszczak2014NJP,Rubio2019NJP,DemkowiczDobrzanski2020JPAMT,Sidhu2020AQS,Rubio2024}. Crucially, the $\{\Pi_x\}$ and $\hat \theta(x)$ thus chosen are parameter-independent, implying that global estimations are operationally meaningful even if little is known \textit{a priori} about $\theta$. Note that the first term on the right hand side of Eq.~(\ref{eq:global_inequality}) is fixed once $p(\theta)$ is given. By contrast, the second term $\mathrm{tr}(\bar\rho S^2)$ depends on $\rho_\theta$ and represents the information about $\theta$ gained in a global estimation. We denote 
\begin{equation}\label{J}
    \mathcal{J}= \mathrm{tr}(\bar\rho S^2).
\end{equation}
To ascertain the ultimate precision achievable for each of the four types of strategies $(i)$-$(iv)$, we need to maximize $\mathcal{J}$ over all the allowed freedom such as the initial state and adaptive controls (see Fig.~\ref{fig:schematic_figure}). However, it is formidable to do so directly  due to the lack of effective tools \cite{Bavaresco2023}.

\textit{Virtual imaginary time evolution}.---To overcome this difficulty, our idea is to establish an equality between the information $\mathcal{J}$ gained in a global estimation and the QFI for a virtual local estimation so that the above maximum can be figured out indirectly by calculating the QFI.

\textit{We first specify the state $\rho_\theta$.} Let $\mathcal{H}_{I_k}$ and $\mathcal{H}_{O_k}$ be the input and output Hilbert spaces of the $k$th copy of the channel $\mathcal{E}_\theta$. Denote by $\mathcal{L}(\mathcal{H})$ the set of linear operators over a Hilbert space $\mathcal{H}$. The $k$th copy of $\mathcal{E}_\theta$, as a completely positive map from $\mathcal{L}(\mathcal{H}_{I_k})$ to $\mathcal{L}(\mathcal{H}_{O_k})$, can be described by a positive semidefinite operator in $\mathcal{L}(\mathcal{H}_{I_k}\otimes\mathcal{H}_{O_k})$,  $E_\theta\coloneqq{id}\otimes\mathcal{E}_\theta\left(| I \rangle\hspace{-2.7pt}\rangle\langle\hspace{-2.7pt}\langle I|\right)$, known as the Choi-Jamio\l{}kowski (CJ) operator \cite{jamiolkowskiLinearTransformationsWhich1972,choiCompletelyPositiveLinear1975}. Here $id$ is the identity map and $| I \rangle\hspace{-2.7pt}\rangle=\sum_j\ket{j}\ket{j}$. The CJ operator of $N$ identical channels is
$C_\theta\coloneqq E_\theta^{\otimes N}\in\mathcal{L}(\mathcal{H}_{I_1}\otimes \mathcal{H}_{O_1}\otimes\cdots\otimes\mathcal{H}_{I_N}\otimes \mathcal{H}_{O_N})$. A strategy is an arrangement of physical operations which, when concatenated with these $N$ channels, produces the output state $\rho_\theta$ carrying information about $\theta$ \cite{liuOptimalStrategiesQuantum2023} (see the left column of Fig.~\ref{fig:schematic_figure}). 
We can therefore regard a strategy as a supermap \cite{Chiribella2008}, taking the $N$ channels as its input and outputting the state $\rho_\theta$ (see the right column of Fig.~\ref{fig:schematic_figure}). This supermap, akin to completely positive maps, can be described by a positive semidefinite operator $X$ in $\mathcal{L}(\mathcal{H}_{I_1}\otimes \mathcal{H}_{O_1}\otimes\cdots\otimes\mathcal{H}_{I_N}\otimes \mathcal{H}_{O_N}\otimes \mathcal{H}_F)$ \cite{liuOptimalStrategiesQuantum2023}. Here $\mathcal{H}_F$ is the Hilbert space upon which $\rho_\theta$ acts \cite{Araujo2017Q}. We have  \cite{chiribellaTheoreticalFrameworkQuantum2009}
\begin{eqnarray}\label{rho}
\rho_\theta=X\star C_\theta.
\end{eqnarray}
Here $\star$ denotes the link product \cite{chiribellaTheoreticalFrameworkQuantum2009}, that is, $X\star C_\theta=\tr_{I_1O_1\cdots I_NO_N}[X(C_\theta^T\otimes\mathbb{I}_F)]$, where $T$ indicates the transpose operation and $\mathbb{I}_F$ represents the identity operator on $\mathcal{H}_F$.

\textit{We next introduce a virtual ITE.} Recall that, whereas traditional time evolution is described by the operator $e^{-iHt}$ with a Hamiltonian $H$, ITE is described by the operator $e^{-H\tau}$ \cite{Motta2019NP}, that is, for a system undergoing ITE, its initial state $\rho(0)$ is evolved to be the state $\rho(\tau)=e^{-H\tau}\rho(0)e^{-H\tau}$ at time $\tau$.
It is interesting to note that the expression $e^{-H\tau}$ is obtained from the expression $e^{-iHt}$ by setting $t$ to be purely imaginary, i.e., $t=-i\tau$ with $\tau\in\mathbb R$. It is also interesting to note that ITE, as a map, is completely positive but not trace-preserving because of its non-unitary character. With the above knowledge, we introduce the averaged CJ operator
$\bar{C}\coloneqq \int{d\theta}p\left( \theta \right) C_{\theta}$ and assume that $\bar{C}$ is evolved to be
\begin{equation}\label{ITE1}
    \bar{C}(\tau) \coloneqq e^{-H\tau}\bar{C}e^{-H\tau}
\end{equation}
at time $\tau$.
Here $H$ is a Hermitian operator in $\mathcal{L}(\mathcal{H}_{I_1}\otimes \mathcal{H}_{O_1}\otimes\cdots\otimes\mathcal{H}_{I_N}\otimes \mathcal{H}_{O_N})$ that satisfies
\begin{equation}\label{ITE2}
    \overline{\theta C}+\left\{ H,\bar{C} \right\} =0,
\end{equation}
with $\overline{\theta C}\coloneqq \int{d\theta}p\left( \theta \right) \theta C_\theta $.
Equations (\ref{ITE1}) and (\ref{ITE2}) specify the ITE of interest here. We clarify that the physical realization of this ITE is irrelevant in the present work, because our purpose is to devise an effective approach to computing $\mathcal{J}$. Hence we regard the ITE as a virtual process. We employ it to define the following family of states, 
\begin{equation}\label{sigma}
\sigma_\tau \coloneqq X\star\bar{C}(\tau).
\end{equation}
Note that $\sigma_\tau$ is positive semidefinite but its trace may not equal to $1$ for $\tau\neq 0$. 

\textit{We now establish an equality between $\mathcal{J}$ and QFI.}
To serve our purpose, we still define the QFI for $\sigma_\tau$ via Eqs.~(\ref{QFI1}) and (\ref{QFI2}) although $\tr(\sigma_\tau)\neq 1$ when $\tau\neq 0$. That is, 
\begin{eqnarray}\label{QFI1-sigma}
	\mathcal{I}[\sigma_\tau]=\mathrm{tr}\left(\sigma_\tau L_\tau^2 \right),
\end{eqnarray}
where $L_\tau$ is the Hermitian operator satisfying $\frac{d}{d\tau}\sigma_\tau =\left( \sigma_\tau L_\tau +L_\tau \sigma _\tau \right)/2$. Using Eqs.~(\ref{rho}), (\ref{ITE1}), (\ref{ITE2}), and (\ref{sigma}), we have
\begin{equation}\label{eq:def_rhobar_thetarhobar}
\sigma _\tau|_{\tau =0}=\bar{\rho},~~~~
\frac{d}{d\tau} \sigma _\tau |_{\tau =0}=\overline{\theta \rho }.
\end{equation}
Further, noting that inserting Eq.~(\ref{eq:def_rhobar_thetarhobar}) into Eq.~(\ref{S}) yields the equality $S=L_\tau |_{\tau=0}$, we arrive at the following theorem.
\begin{theorem} \label{theorem1}
Let $\mathcal{I}[\sigma_\tau]$ be the QFI defined by Eq.~(\ref{QFI1-sigma}). Then 
\begin{eqnarray}\label{J-QFI}
\mathcal{J}=\mathcal{I}[\sigma_\tau]|_{\tau=0},
\end{eqnarray}
i.e., the information $\mathcal{J}$ gained in the global estimation with $\rho_\theta$ is equal to the QFI $\mathcal{I}[\sigma_\tau]$ for the local estimation with $\sigma_\tau$ at $\tau=0$.
\end{theorem}
\noindent Thus we can alternatively compute $\mathcal{I}[\sigma_\tau]$ for obtaining  $\mathcal{J}$. We emphasize that, despite $\tr(\sigma_\tau)\neq 1$, it is still possible to use existing tools to compute $\mathcal{I}[\sigma_\tau]$. To illustrate this point and also for later usage, we prove in Supplemental Material (SM) \cite{SM} that the formula proposed by Fujiwara and Imai \cite{fujiwaraFibreBundleManifolds2008} remains applicable when $\tr(\sigma_\tau)\neq 1$.
\begin{theorem}[Fujiwara and Imai's formula] \label{theorem2}
Let $\left\{ |\psi _j\rangle \right\}_{j=1}^q$ be an ensemble of pure states for $\sigma _\tau $, namely, $\sigma _\tau =\sum_{j=1}^q{\ket{\psi _j}\bra{\psi _j}}$, where $q\ge \mathrm{rank}(\sigma _\tau) $ is an integer. Then
\begin{equation}
\mathcal{I}[\sigma_\tau]|_{\tau=0}=\underset{\left\{ |\psi _j\rangle \right\}_{j=1}^q}{\min}\,\,4\left. \mathrm{tr}\left( \sum_{j=1}^q\ket{\dot{\psi_j}}\bra{ \dot{\psi_j}} \right) \right|_{\tau =0},
\end{equation}
where $\ket{\dot{\psi_j}} =d|\psi _j\rangle /d\tau $ and the minimum is taken over all the ensembles with fixed $q$.
\end{theorem}

\textit{Maximal information gained in global estimation strategies.}---We use index $k$ to specify the type of strategies in question, that is, $k=i, ii, iii, iv$ refer to the four types of strategies ($i$)-($iv$), respectively. Note that
the value of $\mathcal{J}$ depends on the specific strategy adopted. We are interested in the maximum of  $\mathcal{J}$ over all the strategies of type $k$. Hereafter we denote this maximum by $\mathcal{J}_\textrm{max}^{(k)}$. Besides we define $\mathbb{X}^{(k)}$ to be the collection of $X$ that can describe all the strategies of type $k$. Using Theorem \ref{theorem1} as well as substituting Eq.~(\ref{sigma}) into Eq.~(\ref{J-QFI}), we can write  $\mathcal{J}_\textrm{max}^{(k)}$ as 
\begin{eqnarray}\label{def-J-max}
\mathcal{J}_\textrm{max}^{(k)}=\max_{{X}\in\mathbb{X}^{(k)}}\mathcal{I}[X\star\bar{C}(\tau)]|_{\tau=0},
\end{eqnarray}
which is the maximal information gained in the strategies of type $k$. To ascertain the ultimate precision achievable in the strategies of type $k$, which is $\int d\theta p(\theta)\theta^2-\mathcal{J}_\textrm{max}^{(k)}$ according to Eq.~(\ref{eq:global_inequality}), we need to figure out $\mathcal{J}_\textrm{max}^{(k)}$.
To this end, we write the ensemble decomposition of $\bar{C}$ as $\bar{C}=\sum_{j=1}^q\ket{\phi_j}\bra{\phi_j}=\Phi\Phi^\dagger$ with $\Phi\coloneqq [\ket{\phi_1},\cdots,\ket{\phi_q}]$. We introduce the set $\tilde{\mathbb{X}}^{(k)}\coloneqq\{\tilde{X}=\tr_FX | X\in\mathbb{X}^{(k)}\}$. Using Theorem \ref{theorem2}, we show \cite{SM} that 
\begin{eqnarray}\label{J-minmax}
\mathcal{J}_{\max}^{(k)}=\max_{\tilde{X}\in\tilde{\mathbb{X}}^{(k)}}\min_{h\in\mathbb{H}^{(q)}}\tr[\tilde{X}\Omega(h)],
\end{eqnarray}
representing a computation-friendly formula for $\mathcal{J}_{\max}^{(k)}$.
Here $\mathbb{H}^{(q)}$ denotes the set of all $q\times q$ Hermitian matrices and 
\begin{eqnarray}
\Omega(h)=4(H^*\Phi^*-i\Phi^* h)(H^*\Phi^*-i\Phi^* h)^\dagger.
\end{eqnarray}
Notably, formula (\ref{J-minmax}), which is first derived here for global estimations, is analogous to those for local estimations \cite{altherrQuantumMetrologyNonMarkovian2021,liuOptimalStrategiesQuantum2023}.

\textit{Semidefinite programs for computing $\mathcal{J}_{\max}^{(k)}$.}---We now convert Eq.~(\ref{J-minmax}) into two semidefinite programs (SDPs) for computing $\mathcal{J}_{\max}^{(k)}$. To do this, we resort to the process matrix formalism \cite{araujoWitnessingCausalNonseparability2015}, which allows us to characterize $\tilde{\mathbb{X}}^{(k)}$ as
\begin{eqnarray}
\tilde{\mathbb{X}}^{(k)}=\{\tilde{X}|\tilde{X}\geq 0,~ \Lambda^{(k)}(\tilde{X})=\tilde{X}, ~\tr\tilde{X}=d_O\},
\end{eqnarray}
when $k=i, ii, iv$. Here, $d_O=\textrm{dim}(\mathcal{H}_{O_1}\otimes\cdots\otimes\mathcal{H}_{O_N})$ and $\Lambda^{(k)}$ is a linear map whose expression can be found in SM \cite{SM} (see also Ref.~\cite{araujoWitnessingCausalNonseparability2015}). The discussion on strategies $(iii)$ needs to be carried out separately and is left to SM \cite{SM}.
We show that $\mathcal{J}_{\max}^{(k)}$ can be computed via the SDP \cite{SM}
\begin{equation} \label{primal_SDP}
	\begin{aligned}
		\underset{\tilde{X},B,C}{\max}\quad&-\mathrm{tr}C-4\,\mathrm{Re} [\mathrm{tr}\left( H^* \Phi ^*B\right)] 
		\\
		\mathrm{s}.\mathrm{t}.\quad&\Lambda^{({k})}(\tilde{X}) =\tilde{X},\quad\mathrm{tr}\tilde{X}=d_O,
		\\
		&\left[ \begin{matrix}
			\tilde{X}&		B^{\dagger}\\
			B&		C\\
		\end{matrix} \right] \ge 0,
		\\
		&B\Phi ^*~\mathrm{is}~\mathrm{Hermitian},
	\end{aligned}
\end{equation}
referred to as the primal SDP. 
As demonstrated below, the gaps among $\mathcal{J}_{\max}^{(k)}$'s may be small, due to which numerical errors in the SDP may compromise the reliability of computed results.
To overcome this issue, we propose Algorithm 1 in SM \cite{SM}. Using Algorithm 1 to assist the primal SDP, we can obtain reliable lower bounds on $\mathcal{J}_{\max}^{(k)}$, for which numerical errors are eliminated.
Further, to obtain upper bounds, we derive the dual SDP
\begin{equation} \label{dual_SDP}
	\begin{aligned}
		\underset{\tilde{Y},\lambda ,h}{\min}\quad&\lambda 
		\\
		\mathrm{s}.\mathrm{t}.\quad&\Lambda^{(k)} \left( \tilde{Y} \right) =0,		
		\\
		&\left[ \begin{matrix}
			 \frac{\lambda}{d_O}+\tilde{Y}&		2\left( H^*\Phi ^*-i\Phi ^*h \right)\\
			2\left( H^*\Phi ^*-i\Phi ^*h \right) ^{\dagger}&		\mathbb{I}_{q}\\
		\end{matrix} \right] \ge 0,
		\end{aligned}
\end{equation}
where $\mathbb{I}_{q}$ is the $q\times q$ identity matrix \cite{SM}. 
Likewise, Algorithm 2 is proposed to eliminate numerical errors in the dual SDP \cite{SM}. Notably, compared with those in Refs.~\cite{altherrQuantumMetrologyNonMarkovian2021,liuOptimalStrategiesQuantum2023}, our SDPs feature significantly fewer constraints. This enables our algorithms to yield tight bounds.
 
\textit{Strict hierarchy}.---Clearly, the four types of strategies ($i$)-($iv$) form a hierarchy $\mathcal{J}_{\max}^{(i)}\leq\mathcal{J}_{\max}^{(ii)}\leq\mathcal{J}_{\max}^{(iii)}\leq\mathcal{J}_{\max}^{(iv)}$, since each of them is a superset of the preceding one. We show that all the three inequalities can be strictly satisfied simultaneously. To this end, we examine the channel
\begin{eqnarray}\label{Example-Channel}
\mathcal{E}_\theta=\mathcal{E}^{(\textrm{AD})}\circ\mathcal{E}^{(\textrm{BF})}\circ\mathcal{U}_\theta,
\end{eqnarray}
composed of the unitary channel $\mathcal{U}_\theta$ with the Kraus operator $e^{-i\theta \sigma_z/2}$, the bit flip channel $\mathcal{E}^{(\textrm{BF})}$ with the Kraus operators $K_{1}^{( \mathrm{BF})}=\sqrt{\eta}\mathbb{I}$ and $K_{2}^{( \mathrm{BF})}=\sqrt{1-\eta}\sigma _x$, and the amplitude damping channel $\mathcal{E}^{(\textrm{AD})}$ with the Kraus operators 
\begin{equation}
	K_{1}^{\left( \mathrm{AD} \right)}=\left[ \begin{matrix}
		1&		0\\
		0&		\sqrt{1-\gamma}\\
	\end{matrix} \right],~~K_{2}^{\left( \mathrm{AD} \right)}=\left[ \begin{matrix}
		0&		\sqrt{\gamma}\\
		0&		0\\
	\end{matrix} \right],
\end{equation}
where $\sigma_\alpha$, $\alpha=x, y, z$, denote the Pauli matrices. Hereafter, we set $p(\theta)$ to be the uniform probability distribution, that is, $p(\theta)=1/(2\pi)$ for $\theta\in[-\pi,\pi)$. Applying the two SDPs as well as the two algorithms to the channel in Eq.~(\ref{Example-Channel}) with $\eta=1/2$ and $\gamma=7/10$, we can show that $\mathcal{J}_{\max}^{(i)}\leq 0.5516<0.5572\leq\mathcal{J}_{\max}^{(ii)}\leq0.5574<0.5703\leq \mathcal{J}_{\max}^{(iii)}\leq 0.5705<0.57053\leq\mathcal{J}_{\max}^{(iv)}$ when $N=2$. We therefore reach the following theorem.
\begin{theorem} \label{theorem3}
There exist parameter estimation problems for which 
\begin{equation} \label{eq:strict_hierarchy}
\mathcal{J}_{\max}^{(i)}<\mathcal{J}_{\max}^{(ii)}<\mathcal{J}_{\max}^{(iii)}<\mathcal{J}_{\max}^{(iv)},
\end{equation}
i.e., the strict hierarchy of ultimate precision can hold for global estimation strategies ($i$)-($iv$).
\end{theorem}

We point out that the hierarchy phenomenon reported in Theorem \ref{theorem3} is not exclusive to the above specific example. We have randomly generated 1000 channels and found that 780 of them obey the strict hierarchy \cite{SM}.

\textit{Unexpected results.}---To illuminate distinct features of global estimations, we make reference to the work by Giovannetti \textit{et al.}~\cite{Giovannetti2006PRL}, where parallel and sequential strategies are examined within the local estimation framework. We need to consider the unitary channel $\mathcal{E}_\theta=\mathcal{U}_\theta$, i.e., the channel in Eq.~(\ref{Example-Channel}) with $\eta=1$ and $\gamma=0$. Hereafter, we refer to the parallel strategies examined within the local (global) estimation framework as local (global) parallel strategies, and similarly for local (global) sequential strategies.
It is well-known that the optimal input state in local parallel strategies is the Greenberger–Horne–Zeilinger (GHZ) state $\left(\ket{0}^{\otimes N}+\ket{1}^{\otimes N}\right)/\sqrt{2}$ for all $\theta\in[-\pi,\pi)$ \cite{Giovannetti2006PRL}. As such, an intuitive deduction may be that the optimal probe state would also be the GHZ state for global parallel strategies. However, we find that $1.5217\leq\mathcal{J}_{\max}^{(i)}\leq 1.5218$ but $\mathcal{J}=1/4$ for the GHZ state when $N=2$, implying that the above deduction is invalid in general. Also, our result suggests that one candidate for the optimal input state in global parallel strategies is $\sqrt{3/10}|0000\rangle +\sqrt{1/5}|0101\rangle +\sqrt{1/5}|1010\rangle +\sqrt{3/10}|1111\rangle$, indicating that the optimal input state differs in structure from the GHZ state. We next switch our discussion to sequential strategies. Recall that adaptive controls are useless in improving the ultimate performance of local sequential strategies \cite{Giovannetti2006PRL}. However, we find that adaptive controls are useful for global sequential strategies. Indeed, when $N=2$, the maximal information attained in global sequential strategies without controls is $0.25$, which is strictly less than the maximal information attained with controls $1.5217\leq\mathcal{J}_{\max}^{(ii)}\leq 1.5219$. Finally, in contrast to the result that local parallel and sequential strategies share the same ultimate performance for every $N$ and every $\theta\in[-\pi,\pi)$ \cite{Giovannetti2006PRL}, we find that global sequential strategies can be superior to global parallel strategies in ultimate performance. We illustrate this point by showing that 
 $\mathcal{J}_{\max}^{(i)}\leq 1.84507<1.84517\leq \mathcal{J}_{\max}^{(ii)}$ when $N=3$.
 
Besides we present in SM \cite{SM} more numerical results on different priors. 

\textit{Concluding remarks}.---A bottleneck hindering the current research in global estimations is the lack of effective tools, which is in sharp contrast to the situation that many such tools are available in local estimations. In this Letter, we have advocated a pathway to surmount this bottleneck. 	
The key innovation here is the technique of the virtual ITE, in which the fictitious state $\sigma_\tau$ is constructed such that its QFI equals to $\mathcal{J}$ at $\tau=0$, its dependence on $X$ is linear, and its ensemble decompositions are easy to find. This opens up the exciting possibility of solving crucial problems in the realm of global estimations by leveraging powerful tools tailored for local estimations. We have demonstrated this possibility by successfully solving the hierarchy problem in global estimations with Fujiwara and Imai's formula, a useful tool in local estimations. Meanwhile, we have uncovered a number of unexpected results, highlighting that a same quantum resource may assume disparate roles when scrutinized through the lenses of local and global estimations. For instance, while the GHZ state is optimal in local estimations, its utility diminishes in global estimations. These captivating differences underscore the need for further explorations of global estimations. Lastly, we remark that our technique makes it possible to construct optimal strategies in global estimations \cite{Bisio2011PRA,Wechs2021PQ} and devise tight bounds in the non-asymptotic regime, which are two research directions for future studies.

The codes used in this article are openly available from \cite{githubbayeoptimal}.

\let\oldaddcontentsline\addcontentsline% Store \addcontentsline
\renewcommand{\addcontentsline}[3]{}% Make \addcontentsline a no-op
\begin{acknowledgments}
We thank Jes\'{u}s Rubio for helpful comments. This work was supported by the National Natural Science Foundation of China through Grant Nos.~12275155 and 12174224.
\end{acknowledgments}

%\bibliography{ref.bib}

\let\addcontentsline\oldaddcontentsline% Restore \addcontentsline
% https://tex.stackexchange.com/questions/345569/remove-bibliography-part-in-table-of-contents

% Supplemental Material
\onecolumngrid
\setcounter{equation}{0}
\renewcommand{\theequation}{S\arabic{equation}}
\renewcommand{\thefigure}{S\arabic{figure}}
\renewcommand{\thetheorem}{S\arabic{theorem}}
\renewcommand{\thesection}{S\arabic{section}}

\def\thA{{2}}
\def\thH{{3}}
\def\eqgsld{{4}}
\def\eqsigmataudtau{{11}}
\def\eqJmaxminopt{{15}}
\def\eqpsdp{{18}}
\def\eqdsdp{{19}}

\setcounter{section}{0}
\setcounter{equation}{0}
\setcounter{figure}{0}
\setcounter{table}{0}
\setcounter{theorem}{0}
\setcounter{lemma}{0}
\clearpage
\onecolumngrid
\begin{center}
	\Large
	\textbf{Supplemental Material for\\ ``Strict hierarchy of optimal strategies for global estimations: Linking global estimations with local ones''}
\end{center}
\vspace{\baselineskip}
\vspace{\baselineskip}
\vspace{\baselineskip}
\vspace{\baselineskip}
\tableofcontents
\vspace{\baselineskip} % Adds space equivalent to one line of text
\vspace{\baselineskip}
\vspace{\baselineskip}
\vspace{\baselineskip}
\vspace{\baselineskip}

\section{Proof of Theorem $\thA$}

Let us first specify some notations and facts. In the following, we refer to an ensemble of $\sigma_\tau$ that consists of $q$ pure states $\ket{\psi_j}$, $j=1,\cdots,q$, as a $\sigma_\tau$-ensemble of size $q$. To compactly describe such an ensemble, we adopt the notation
\begin{equation}
    W=\left[\ket{\psi_1},\ket{\psi_2},\cdots,\ket{\psi_q}\right].
\end{equation}
Let the spectral decomposition of $\sigma_\tau$ be $\sigma _{\tau}=\sum_{j=1}^r{|\varphi_j\rangle}\langle \varphi_j|$ with $r=\textrm{rank}(\sigma_\tau)$. Clearly, $\{\ket{\varphi_j}\}_{j=1}^r$ is a $\sigma_\tau$-ensemble of size $r$. We use
\begin{equation}
    W^{(0)}=\left[\ket{\varphi_1},\ket{\varphi_2},\cdots,\ket{\varphi_r}\right]
\end{equation}
to describe this ensemble. It is not difficult to see that $W^{(0)\dagger}W^{(0)}$ is an invertible matrix. Noting that two ensembles of $\sigma_\tau$ can be transformed into each other under a unitary matrix $U$ (see the theorem on page 103 of Ref.~\cite{Nielsen2012}), we have
\begin{equation}\label{W-U}
    W=W^{(0)}V^{(0)}U.
\end{equation}
Here, $V^{(0)}\coloneqq [\mathbb{I}|\mathbb{O}]$, where $\mathbb{I}$ and $\mathbb{O}$ denote the $r\times r$ identity matrix and the $r\times(q-r)$ zero matrix. The multiplication $W^{(0)}V^{(0)}$ amounts to padding the ensemble $\{\ket{\varphi_j}\}_{j=1}^r$ with $(q-r)$ additional vectors $0$. We clarify that $U$ may depend on $\tau$. Note that any $\sigma_\tau$-ensemble of size $q$ can be expressed in the form (\ref{W-U}), where $W^{(0)}V^{(0)}$ is the reference ensemble and $U$ can be an arbitrary $q\times q$ unitary matrix. Using the equality $\dot{W}\dot{W}^\dagger=\sum_{j=1}^q\ket{\dot{\psi}_j}\bra{\dot{\psi}_j}$, we can rewrite
\begin{equation}
    \left .\mathcal{I}[\sigma_\tau]\right|_{\tau=0}=\underset{\left\{ |\psi _j\rangle \right\} _{j=1}^{q}}{\min}\,\,4\left. \mathrm{tr}\left( \sum_{j=1}^q{|\dot{\psi}_j\rangle}\langle \dot{\psi}_j| \right) \right|_{\tau =0}
\end{equation}
as
\begin{equation} \label{app_eq_Wexpression_min_ensemble}
    \left .\mathcal{I}[\sigma_\tau]\right|_{\tau=0}=\underset{W}{\min}\,\,4\left. \mathrm{tr}\left( \dot{W}\dot{W}^{\dagger} \right) \right|_{\tau =0},
\end{equation}
where the minimum is taken over all of the $\sigma_\tau$-ensembles of size $q$. To prove Theorem \thA, we need the following lemma.

\begin{lemma}\label{lemma1}
    The following two statements are equivalent:\\
    (a) There exists a unitary matrix $U^{(\mathrm{op})}$ and a Hermitian matrix $L_\tau$ such that $W^{(\mathrm{op})}=W^{(0)}V^{(0)}U^{(\mathrm{op})}$ satisfies
    \begin{equation} \label{app_eq_dotWLW}
        \dot W^{(\mathrm{op})}=\frac{1}{2}L_\tau W^{(\mathrm{op})}
    \end{equation}
    at $\tau=0$. \\
    (b) There exists a Hermitian matrix $L_\tau$ satisfying
    \begin{equation} \label{app_eq_SLD_sigmatau}
        \frac{d}{d\tau}\sigma _{\tau}=\frac{1}{2}\left( \sigma _{\tau}L_{\tau}+L_{\tau}\sigma _{\tau} \right)
    \end{equation}
    at $\tau=0$.
\end{lemma}

\textit{Proof.} That (a) implies (b) can be confirmed by using the equality  $\sigma_\tau=W^{(\mathrm{op})}W^{(\mathrm{op})\dagger}$ and substituting \cref{app_eq_dotWLW} into \cref{app_eq_SLD_sigmatau}.

To show that (b) implies (a), we make use of the following ansatz
\begin{equation} \label{app_eq_def_Uop}
    U^{(\mathrm{op})}=e^{iD\tau},
\end{equation}
where
\begin{equation} \label{app_eq_defD}
    D=V^{(0)\dagger}\tilde DV^{(0)},
\end{equation}
with
\begin{equation} \label{app_eq_def_tildeD}
    \tilde{D}=i\left.\left( \frac{1}{2}W^{\left( 0 \right) \dagger}L_{\tau}-\dot{W}^{\left( 0 \right) \dagger} \right) W^{\left( 0 \right)}\left( W^{\left( 0 \right) \dagger}W^{\left( 0 \right)} \right) ^{-1} \right|_{\tau=0}.
\end{equation}
We need to prove that the $U^{(\mathrm{op})}$ in \cref{app_eq_def_Uop}
is unitary and the associated
$W^{(\mathrm{op})}=W^{(0)}V^{(0)}U^{(\mathrm{op})}$ satisfies \cref{app_eq_dotWLW} at $\tau=0$.

To prove that the $U^{(\mathrm{op})}$ given by \cref{app_eq_def_Uop}
is unitary, we need to show that $D$ is Hermitian, which holds if and only if $\tilde{D}$ is Hermitian, as can be easily seen from \cref{app_eq_defD}. Using \cref{app_eq_def_tildeD}
and noting that $\sigma_\tau=W^{(0)}W^{(0)\dagger}$, we have that
\begin{equation}\label{D}
    \left(W^{(0)\dagger}W^{(0)}\right)\tilde{D}\left(W^{(0)\dagger}W^{(0)}\right)=iW^{(0)\dagger}\left(\frac{1}{2}\sigma_\tau L_\tau-W^{(0)}\dot{W}^{(0)\dagger}\right)W^{(0)}.
\end{equation}
Likewise, we have, after some algebra,
\begin{equation}\label{D-dagger}
    \left(W^{(0)\dagger}W^{(0)}\right)\tilde{D}^
    \dagger\left(W^{(0)\dagger}W^{(0)}\right)=-iW^{(0)\dagger}\left(\frac{1}{2} L_\tau\sigma_\tau-\dot{W}^{(0)}{W}^{(0)\dagger}\right)W^{(0)}.
\end{equation}
Further, noting that \cref{app_eq_SLD_sigmatau} implies that
\begin{equation}\label{alter-s7}
    \frac{1}{2}\sigma_\tau L_\tau-W^{(0)}\dot{W}^{(0)\dagger}=-\left(\frac{1}{2} L_\tau\sigma_\tau-\dot{W}^{(0)}{W}^{(0)\dagger}\right),
\end{equation}
we deduce from Eqs.~(\ref{D}) and (\ref{D-dagger})
that
\begin{equation}
    \left(W^{(0)\dagger}W^{(0)}\right)\tilde{D}\left(W^{(0)\dagger}W^{(0)}\right)=\left(W^{(0)\dagger}W^{(0)}\right)\tilde{D}^
    \dagger\left(W^{(0)\dagger}W^{(0)}\right).
\end{equation}
Then, as $\left(W^{(0)\dagger}W^{(0)}\right)$ is invertible, we have $\tilde{D}=\tilde{D}^\dagger$, that is, $\tilde{D}$ is Hermitian.

To prove that the $W^{(\mathrm{op})}=W^{(0)}V^{(0)}U^{(\mathrm{op})}$ with the $U^{(\mathrm{op})}$ in \cref{app_eq_def_Uop} satisfies \cref{app_eq_dotWLW} at $\tau=0$, we need to show that
\begin{equation}\label{alter-s6}
    \dot{W}^{\left( 0 \right)}V^{\left( 0 \right)}+iW^{\left( 0 \right)}V^{\left( 0 \right)}D=\frac{1}{2}L_{\tau}W^{\left( 0 \right)}V^{\left( 0 \right)}.
\end{equation}
Using \cref{app_eq_defD} and noting that $V^{(0)}V^{(0)\dagger}=\mathbb{I}$, we have that the left-hand side of Eq.~(\ref{alter-s6}) reads
\begin{equation}\label{W-V}
    \dot{W}^{\left( 0 \right)}V^{\left( 0 \right)}+iW^{\left( 0 \right)}V^{\left( 0 \right)}D=\left( \dot{W}^{\left( 0 \right)}+iW^{\left( 0 \right)}\tilde{D} \right) V^{\left( 0 \right)}.
\end{equation}
Using \cref{app_eq_def_tildeD}, we have, after some algebra,
\begin{equation}
    \dot{W}^{\left( 0 \right)}+iW^{\left( 0 \right)}\tilde{D}=    \dot{W}^{\left( 0 \right)}-\left( \frac{1}{2}\sigma _{\tau}L_{\tau}-W^{\left( 0 \right)}\dot{W}^{\left( 0 \right) \dagger} \right) W^{\left( 0 \right)}\left( W^{\left( 0 \right) \dagger}W^{\left( 0 \right)} \right) ^{-1},
\end{equation}
which, in conjunction with Eq.~(\ref{alter-s7}), gives
\begin{eqnarray}
    \dot{W}^{\left( 0 \right)}+iW^{\left( 0 \right)}\tilde{D}&=&    \dot{W}^{\left( 0 \right)}+\left( \frac{1}{2}L_{\tau}\sigma _{\tau}-\dot{W}^{\left( 0 \right)}{W}^{\left( 0 \right) \dagger} \right) W^{\left( 0 \right)}\left( W^{\left( 0 \right) \dagger}W^{\left( 0 \right)} \right) ^{-1}\\
    &=&\frac{1}{2}L_\tau\sigma_\tau W^{\left( 0 \right)}\left( W^{\left( 0 \right) \dagger}W^{\left( 0 \right)} \right) ^{-1}\\
    &=&\frac{1}{2}L_\tau W^{(0)}W^{(0)\dagger} W^{\left( 0 \right)}\left( W^{\left( 0 \right) \dagger}W^{\left( 0 \right)} \right) ^{-1}\\
    &=&\frac{1}{2}L_\tau W^{(0)}.\label{L-W}
\end{eqnarray}
Substituting Eq.~(\ref{L-W}) into Eq.~(\ref{W-V}), we obtain Eq.~(\ref{alter-s6}). $\hfill\square$

We now prove Theorem $\thA$. Using Eq.~($\eqsigmataudtau$) in the main text, i.e., $\sigma _\tau|_{\tau =0}=\bar{\rho}$ and $\frac{d}{d\tau} \sigma _\tau |_{\tau =0}=\overline{\theta \rho }$, we can rewrite Eq.~(\ref{app_eq_SLD_sigmatau}) as
\begin{equation} \label{app:lemma3_lyapunov}
    \overline{\theta \rho }=\frac{1}{2}\left( \bar{\rho}L_\tau+L_\tau\bar{\rho} \right)|_{\tau=0}.
\end{equation}
As Eq.~(\ref{app:lemma3_lyapunov}) is nothing but Eq.~(\eqgsld) in the main text, we have that $L_\tau|_{\tau=0}=S$. We clarify that it is known that there exists a Hermitian matrix $S$ satisfying Eq.~(\eqgsld) in the main text. This implies that the statement $(b)$ in Lemma \ref{lemma1} holds. Consequently, the statement $(a)$ in Lemma \ref{lemma1} holds as well.
Using Eq.~(\ref{app_eq_dotWLW}) and noting that $\sigma_\tau=W^{(\textrm{op})}W^{(\textrm{op})\dagger}$, we have
\begin{equation}\label{QFI-W-OP}
    \left .\mathcal{I}[\sigma_\tau]\right |_{\tau=0}=4\left .\mathrm{tr}\left( \dot{W}^{\left( \mathrm{op} \right)}\dot{W}^{\left( \mathrm{op} \right) \dagger} \right)\right |_{\tau=0}.
\end{equation}
On the other hand, note that any $W$ can be written as $W=W^{\left( \mathrm{op} \right)}V$ with $V$ a unitary matrix. Direct calculations show that
\begin{equation}\label{W-WOP}
    \begin{aligned}
        \mathrm{tr}\left( \dot{W}\dot{W}^{\dagger} \right) & =\mathrm{tr}\left[ \left( \dot{W}^{\left( \mathrm{op} \right)}V+W^{\left( \mathrm{op} \right)}\dot{V} \right) \left( \dot{W}^{\left( \mathrm{op} \right)}V+W^{\left( \mathrm{op} \right)}\dot{V} \right) ^{\dagger} \right]
        \\
                                                           & =\mathrm{tr}\left( \dot{W}^{\left( \mathrm{op} \right)}\dot{W}^{\left( \mathrm{op} \right) \dagger}+\dot{W}^{\left( \mathrm{op} \right)}V\dot{V}^{\dagger}W^{\left( \mathrm{op} \right) \dagger}+W^{\left( \mathrm{op} \right)}\dot{V}V^{\dagger}\dot{W}^{\left( \mathrm{op} \right) \dagger}+W^{\left( \mathrm{op} \right)}\dot{V}\dot{V}^{\dagger}W^{\left( \mathrm{op} \right) \dagger} \right)
        \\
                                                           & =\mathrm{tr}\left( \dot{W}^{\left( \mathrm{op} \right)}\dot{W}^{\left( \mathrm{op} \right) \dagger}+\frac{1}{2}L_\tau{W}^{\left( \mathrm{op} \right)}V\dot{V}^{\dagger}W^{\left( \mathrm{op} \right) \dagger}+\frac{1}{2}W^{\left( \mathrm{op} \right)}\dot{V}V^{\dagger}{W}^{\left( \mathrm{op} \right) \dagger}L_\tau+W^{\left( \mathrm{op} \right)}\dot{V}\dot{V}^{\dagger}W^{\left( \mathrm{op} \right) \dagger} \right)
        \\
                                                           & =\mathrm{tr}\left( \dot{W}^{\left( \mathrm{op} \right)}\dot{W}^{\left( \mathrm{op} \right) \dagger}+W^{\left( \mathrm{op} \right)}\dot{V}\dot{V}^{\dagger}W^{\left( \mathrm{op} \right) \dagger} \right)
        \\
                                                           & \ge \mathrm{tr}\left( \dot{W}^{\left( \mathrm{op} \right)}\dot{W}^{\left( \mathrm{op} \right) \dagger} \right).
    \end{aligned}
\end{equation}
Here, the third equality follows from Eq.~(\ref{app_eq_dotWLW}), the fourth equality follows from the cyclic property of trace and  $V\dot{V}^\dagger+\dot{V}V^\dagger=0$, and the last inequality follows from the fact that $W^{\left( \mathrm{op} \right)}\dot{V}\dot{V}^{\dagger}W^{\left( \mathrm{op} \right) \dagger}$ is a positive semidefinite matrix and hence $\tr\left(W^{\left( \mathrm{op} \right)}\dot{V}\dot{V}^{\dagger}W^{\left( \mathrm{op} \right) \dagger}\right)\geq 0$. Using Eqs.~(\ref{QFI-W-OP}) and (\ref{W-WOP}), we arrive at Eq.~(\ref{app_eq_Wexpression_min_ensemble}). This completes the proof of Theorem $\thA$.

\section{Proof of Eq.~(\eqJmaxminopt)}

Here we present the proof of Eq.~(\eqJmaxminopt), which is analogous to those in Refs.~\cite{altherrQuantumMetrologyNonMarkovian2021,liuOptimalStrategiesQuantum2023}. Note that, for any operator $X$ of rank $r$, we can always purify $X$ by adding an $r$-dimensional ancillary system. This only extends the dimension of the Hilbert space $\mathcal{H}_F$ without changing the causal relations between the $N$ channels $\mathcal{E}_\theta$. Therefore, without loss of generality, we can assume that $X$ is of rank $1$. This allows us to write $X$ as
$X=|X\rangle\langle X|$.
Denote by $\{\ket{\phi_j(\tau)}\}_{j=1}^q$ a $\bar{C}(\tau)$-ensemble of size $q$, that is, $\bar{C}(\tau)=\sum_{j=1}^q\ket{\phi_j(\tau)}\bra{\phi_j(\tau)}$. The equation $\sigma_\tau=X\star \bar{C}(\tau)$ implies that the pure states
\begin{eqnarray}\label{psi-phi}
    \ket{\psi_j}=\bra{\phi_j^*(\tau)}\otimes\mathbb{I}_F\ket{X}
\end{eqnarray}
constitute a $\sigma_\tau$-ensemble of size $q$, where $*$ indicates the complex conjugate. This point can be verified directly as follows:
\begin{eqnarray}\label{verify-sigma}
    \sum_{j=1}^q\ket{\psi_j}\bra{\psi_j}&=&\sum_{j=1}^q \left(\bra{\phi_j^*(\tau)}\otimes\mathbb{I}_F\right)\ket{X}\bra{X}\left(\ket{\phi_j^*(\tau)}\otimes\mathbb{I}_F\right)\nonumber\\
    &=&\sum_{j=1}^q \tr_{IO}\left[\ket{X}\bra{X}\left(\ket{\phi_j^*(\tau)}\bra{\phi_j^*(\tau)}\otimes\mathbb{I}_F\right)\right]\nonumber\\
    &=&\tr_{IO}\left[\ket{X}\bra{X}\left(\bar{C}^T(\tau)\otimes\mathbb{I}_F\right)\right]\nonumber\\
    &=&X\star\bar{C}(\tau)\nonumber\\
    &=&\sigma_\tau,
\end{eqnarray}
where $\tr_{IO}$ is the partial trace over $\mathcal{H}_{I_1}\otimes\mathcal{H}_{O_1}\otimes\cdots\otimes\mathcal{H}_{I_N}\otimes\mathcal{H}_{O_N}$. It follows from Eq.~(\ref{psi-phi}) that we can rewrite the equation
\begin{equation}
    \mathcal{I}[\sigma_\tau]|_{\tau=0}=\underset{\left\{ |\psi _j\rangle \right\}_{j=1}^q}{\min}\,\,4\left. \mathrm{tr}\left( \sum_{j=1}^q\ket{\dot{\psi_j}}\bra{ \dot{\psi_j}} \right) \right|_{\tau =0}
\end{equation}
as
\begin{equation}
    \mathcal{I}[\sigma_\tau]|_{\tau=0}=\underset{\{\ket{\phi_j(\tau)}\}_{j=1}^q}{\min}\,\,4\left. \mathrm{tr}\left[ \left(\bra{\dot{\phi}_j^*(\tau)}\otimes\mathbb{I}_F\right)\ket{X}\bra{X}
        \left(\ket{\dot{\phi}_j^*(\tau)}\otimes\mathbb{I}_F\right)\right] \right|_{\tau =0}.
\end{equation}
Here, the minimum is taken over all the $\bar{C}(\tau)$-ensembles of size $q$, as there is a one-to-one correspondence between the $\bar{C}(\tau)$-ensembles of size $q$ and the $\sigma_\tau$-ensembles of size $q$ [see Eq.~(\ref{psi-phi})]. Using the same reasoning as in Eq.~(\ref{verify-sigma}), we have
\begin{equation}
    \mathcal{I}[\sigma_\tau]|_{\tau=0}=\min_{{\{\ket{\phi_j(\tau)}\}_{j=1}^q}} 4\left.\tr\left[\ket{X}\bra{X}\left(\left(\sum_{j=1}^q\ket{\dot{\phi}_j(\tau)}\bra{\dot{\phi}_j(\tau)}\right)^T\otimes\mathbb{I}_F\right)\right] \right|_{\tau =0},
\end{equation}
which leads to
\begin{equation}\label{QFI-expr1}
    \mathcal{I}[\sigma_\tau]|_{\tau=0}=\min_{{\{\ket{\phi_j(\tau)}\}_{j=1}^q}} 4\left.\tr_{IO}\left[\tilde{X}\left(\sum_{j=1}^q\ket{\dot{\phi}_j(\tau)}\bra{\dot{\phi}_j(\tau)}\right)^T\right]\right|_{\tau =0}
\end{equation}
by noting that $\ket{X}\bra{X}=X$ and $\tilde{X}=\tr_FX$. Further, letting $\Phi(\tau)\coloneqq\left[\ket{\phi_1(\tau)},\cdots,\ket{\phi_q(\tau)}\right]$, we can express Eq.~(\ref{QFI-expr1}) as
\begin{equation}\label{QFI-expr2}
    \mathcal{I}[\sigma_\tau]|_{\tau=0}=\min_{\Phi(\tau)} ~~4\left.\tr_{IO}\left[\tilde{X}\left(\dot{\Phi}(\tau)\dot{\Phi}^\dagger(\tau)\right)^T\right]\right|_{\tau =0}.
\end{equation}
Recall that $\bar{C}(\tau)=e^{-H\tau}\bar{C}e^{-H\tau}$ and $\bar{C}=\Phi\Phi^\dagger$. Here, as defined in the main text, $\Phi$ denotes any fixed $\bar{C}$-ensemble of size $q$. Note that we distinguish between the two notations $\Phi(\tau)$ and $\Phi$. Without loss of generality, we can express $\Phi(\tau)$ as
\begin{eqnarray}\label{Phi-tau}
    \Phi(\tau)=e^{-H\tau}\Phi U,
\end{eqnarray}
where $U$ can be taken to be any $q\times q$ unitary matrix. Substituting Eq.~(\ref{Phi-tau}) into Eq.~(\ref{QFI-expr2}), we have, after some algebra,
\begin{equation}\label{QFI-expr3}
    \mathcal{I}[\sigma_\tau]|_{\tau=0}=\min_{h} ~~4\tr_{IO}\left[\tilde{X}\left(\left(H\Phi+i\Phi h^T\right)\left(H\Phi+i\Phi h^T\right)^\dagger\right)^T\right],
\end{equation}
where we have defined $h^T=i\dot{U}U^\dagger$. Here, as $U$ is arbitrary, the minimum is taken over all the $q\times q$ Hermitian matrices $h$. Finally, using $\Omega(h)$ to represent the term $4\left(\left(H\Phi+i\Phi h^T\right)\left(H\Phi+i\Phi h^T\right)^\dagger\right)^T$, we arrive at Eq.~(\eqJmaxminopt) in the main text.

\section{Characterizations of $\tilde{\mathbb{X}}^{(k)}$}
\label{app_characterization_Xk}

\begin{figure}
    \centering
    \includegraphics[width=\linewidth]{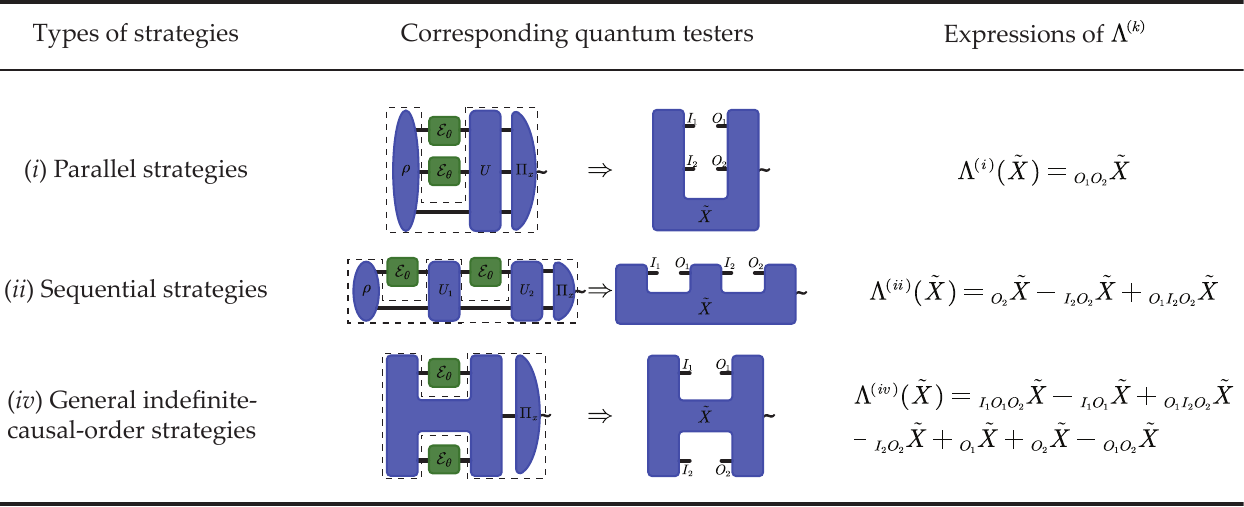}
    \caption{Characterizations of $\tilde{\mathbb{X}}^{(k)}$ for $k=i, ii, iv$ when $N=2$.}
    \label{fig_SM_tester}
\end{figure}

In what follows, we characterize $\tilde{\mathbb{X}}^{(k)}$ for $k=i, ii, iv$, postponing the discussions on the strategies of type $iii$ to Sec.~\ref{app_sec_strategiesiii}. To this end, we resort to the process matrix formalism \cite{araujoWitnessingCausalNonseparability2015}, which has been used to characterize quantum testers \cite{Bavaresco2021}. We denote a POVM by $\{\Pi_x\}_x$ and introduce the operator $T_x\coloneqq X\star\Pi_x^T$. According to the process matrix formalism \cite{chiribellaTheoreticalFrameworkQuantum2009,araujoWitnessingCausalNonseparability2015,Bavaresco2021}, the collection of operators
$\{T_x\}_x$ defines a quantum tester which,
physically speaking, describes the process of concatenating the strategy $X$ with the quantum measurement $\{\Pi_x\}$ (see Fig.~\ref{fig_SM_tester}). Using $\sum_x\Pi_x=\mathbb{I}_F$, we have
\begin{eqnarray}\label{chara-X-tester}
    \tilde{X}=\sum_x T_x,
\end{eqnarray}
i.e., $\tilde{X}=\tr_FX$ is related to the quantum tester $\{T_x\}_x$ via Eq.~(\ref{chara-X-tester}). Note that different types of strategies correspond to different types of quantum testers. For example, parallel, sequential, and general indefinite-causal-order strategies correspond to parallel, sequential, and general testers, respectively (see Ref.~\cite{Bavaresco2021} for the definitions of these testers). Note also that the characterizations of these quantum testers have been given in Refs.~\cite{chiribellaTheoreticalFrameworkQuantum2009,araujoWitnessingCausalNonseparability2015,Bavaresco2021}. From these known results, we can directly obtain the characterizations of $\tilde{\mathbb{X}}^{(k)}$. Specifically, from the result on the characterization of parallel testers (see Eq.~(13) in Ref.~\cite{araujoWitnessingCausalNonseparability2015} and Eq.~(1) in Ref.~\cite{Bavaresco2021}), it follows that $\tilde{\mathbb{X}}^{(i)}$ can be characterized as
\begin{eqnarray}
    \tilde{\mathbb{X}}^{(i)}=\{\tilde{X}|\tilde{X}\geq 0,~~ \Lambda^{(i)}(\tilde{X})=\tilde{X},~ ~\tr\tilde{X}=d_O\},
\end{eqnarray}
where, as mentioned in the main text, $d_O$ is the dimension of $\mathcal{H}_{O_1}\otimes\cdots\otimes\mathcal{H}_{O_N}$ and $\Lambda^{(i)}(\tilde{X})$ is a linear map defined as
\begin{eqnarray}
    \Lambda^{(i)}(\tilde{X})\coloneqq\prescript{}{O_1\cdots O_N}{\tilde{X}}.
\end{eqnarray}
Here and henceforth, the symbol $\prescript{}{Q}{\tilde{X}}$ denotes the completely positive and trace-preserving map consisting of tracing out the subsystem $Q$ and replacing it with the normalized identity operator,
$\prescript{}{Q}{\tilde{X}}\coloneqq \tr_Q\tilde{X}\otimes (\mathbb{I}_Q/d_Q)$, where $d_Q=\textrm{dim}(\mathcal{H}_Q)$ \cite{araujoWitnessingCausalNonseparability2015}.
Likewise, from the result on the characterization of sequential testers (see Eq.~(15) in Ref.~\cite{araujoWitnessingCausalNonseparability2015} and Eqs.~(2) and (3) in Ref.~\cite{Bavaresco2021}), it follows that $\tilde{\mathbb{X}}^{(ii)}$ is the collection of the operators $\tilde{X}$ that satisfy $\tilde{X}\geq 0$, $\tr\tilde{X}=d_O$, and
\begin{eqnarray}\label{chara-seq}
    \begin{aligned}
        \tilde{X}                                    & =\prescript{}{O_N}{\tilde{X}},                    \\
        \prescript{}{I_NO_N}{\tilde{X}}              & =\prescript{}{O_{N-1}I_NO_N}{\tilde{X}},          \\
                                                     & ~~\vdots                                          \\
        \prescript{}{I_2O_2\cdots I_NO_N}{\tilde{X}} & =\prescript{}{O_1I_2O_2\cdots I_NO_N}{\tilde{X}}.
    \end{aligned}
\end{eqnarray}
We show that the above characterization can be equivalently formulated as
\begin{eqnarray}\label{chara-seq-set}
    \tilde{\mathbb{X}}^{(ii)}=\{\tilde{X}|\tilde{X}\geq 0,~~ \Lambda^{(ii)}(\tilde{X})=\tilde{X},~ ~\tr\tilde{X}=d_O\},
\end{eqnarray}
with
\begin{eqnarray}\label{chara-Lambda-Seq}
    \Lambda^{(ii)}(\tilde{X})\coloneqq\prescript{}{O_N}{\tilde{X}}-
    \prescript{}{(1-O_{N-1})I_NO_N}{\tilde{X}}-
    \prescript{}{(1-O_{N-2})I_{N-1}O_{N-1}I_NO_N}{\tilde{X}}-\cdots-
    \prescript{}{(1-O_1)I_2O_2\cdots I_NO_N}{\tilde{X}},
\end{eqnarray}
where $\prescript{}{(1-Q)}{\tilde{X}}\coloneqq\tilde{X}-\prescript{}{Q}{\tilde{X}}$.
To our knowledge, the above equivalence has not been explicitly pointed out in the literature and, therefore, we need to prove it. To do this, we introduce a family of maps, $P_{j}$, $j=1,\cdots,N$, defined as
\begin{eqnarray}
    \begin{aligned}
        P_N(\tilde{X})     & \coloneqq\prescript{}{(1-O_N)}{\tilde{X}},                      \\
        P_{N-1}(\tilde{X}) & \coloneqq\prescript{}{(1-O_{N-1})I_NO_N}{\tilde{X}},            \\
                           & ~~\vdots                                                        \\
        P_{1}(\tilde{X})   & \coloneqq\prescript{}{(1-O_{1})I_2O_2\cdots I_NO_N}{\tilde{X}}.
    \end{aligned}
\end{eqnarray}
A useful property of these maps is that $P_iP_j=\delta_{ij}P_i$, where $\delta_{ij}$ denotes the Kronecker delta. This can be verified by noting that $\prescript{}{O_iO_j}{\tilde{X}}=\prescript{}{O_jO_i}{\tilde{X}}$, $\prescript{}{I_iI_j}{\tilde{X}}=\prescript{}{I_jI_i}{\tilde{X}}$, $\prescript{}{O_iI_j}{\tilde{X}}=\prescript{}{I_jO_i}{\tilde{X}}$, $\prescript{}{O_i^2}{\tilde{X}}=\prescript{}{O_i}{\tilde{X}}$, $\prescript{}{I_i^2}{\tilde{X}}=\prescript{}{I_i}{\tilde{X}}$, and $\prescript{}{(1-O_j)O_j}{\tilde{X}}=0$.
We can rewrite Eq.~(\ref{chara-seq}) as
\begin{eqnarray}\label{chara-P}
    \begin{aligned}
        (1-P_N)(\tilde{X})     & =\tilde{X}, \\
        (1-P_{N-1})(\tilde{X}) & =\tilde{X}, \\
                               & ~~\vdots    \\
        (1-P_{1})(\tilde{X})   & =\tilde{X}.
    \end{aligned}
\end{eqnarray}
As $P_iP_j=\delta_{ij}P_i$ implies $(1-P_i)(1-P_j)=(1-P_j)(1-P_i)$ and $(1-P_j)^2=1-P_j$, $\tilde{X}$ satisfies Eq.~(\ref{chara-P}) if and only if $\tilde{X}$ satisfies
\begin{eqnarray}\label{chara-P-X}
    \prod_{j=1}^N(1-P_j)(\tilde{X})=\tilde{X}.
\end{eqnarray}
Further, noting that the property $P_iP_j=\delta_{ij}P_i$ implies $\prod_{j=1}^N(1-P_j)=1-\sum_{j=1}^NP_j$, we have that Eq.~(\ref{chara-P-X}) can be rewritten as $\Lambda^{(ii)}(\tilde{X})=\tilde{X}$ with $\Lambda^{(ii)}$ given by Eq.~(\ref{chara-Lambda-Seq}). This completes the proof of the equivalence. Finally,
from the result on the characterization of general testers (see Eq.~(B28) in Ref.~\cite{araujoWitnessingCausalNonseparability2015}), it follows that $\tilde{\mathbb{X}}^{(iv)}$ can be characterized as
\begin{eqnarray}
    \tilde{\mathbb{X}}^{(iv)}=\{\tilde{X}|\tilde{X}\geq 0,~~ \Lambda^{(iv)}(\tilde{X})=\tilde{X},~ ~\tr\tilde{X}=d_O\},
\end{eqnarray}
with
\begin{eqnarray}
    \Lambda^{(iv)}(\tilde{X})\coloneqq\prescript{}{\left[1-\prod_{j=1}^N\left(1-O_j+I_jO_j\right)+\prod_{j=1}^NI_jO_j\right]}{\tilde{X}}.
\end{eqnarray}
By now, we have finished the characterizations of $\tilde{\mathbb{X}}^{(k)}$ for $k=i, ii, iv$. Figure \ref{fig_SM_tester}
shows these characterizations for the $N=2$ case.

\section{Proof of Eq.~(\eqpsdp)}
To prove Eq.~(\eqpsdp), we recall some useful knowledge.
Let $\Psi$ be a Hermitian-preserving map and $\mathsf{A},\mathsf{B}$ be Hermitian operators. A semidefinite program (SDP) is a triple $(\Psi,\mathsf{A},\mathsf{B})$ with which the following optimization problem is associated \cite{Watrous2018}:
\begin{equation} \label{app_sdp_primal}
    \begin{aligned}
        \underset{\mathsf{X}}{\max}\quad & \mathrm{tr}\left( \mathsf{A}\mathsf{X} \right)
        \\
        \mathrm{s}.\mathrm{t}.\quad      & \Psi \left( \mathsf{X}  \right) =\mathsf{B},
        \\
                                         & \mathsf{X}\ge 0.
    \end{aligned}
\end{equation}
Here we distinguish the two symbols $\mathsf{X}$ and $X$.
The above problem is known as the primal problem. The associated dual problem is
\begin{equation} \label{app_sdp_dual}
    \begin{aligned}
        \underset{\mathsf{Y}}{\min}\quad & \mathrm{tr}\left( \mathsf{B}\mathsf{Y} \right)
        \\
        \mathrm{s}.\mathrm{t}.\quad      & \Psi^* \left( \mathsf{Y}  \right) \ge\mathsf{A},
        \\
                                         & \mathsf{Y}=\mathsf{Y}^\dagger,
    \end{aligned}
\end{equation}
where $\Psi^*$, known as the adjoint map of $\Psi$, is defined to be the map satisfying $\mathrm{tr}[ \mathsf{X}^{\dagger}\Psi ( \mathsf{Y} ) ] =\mathrm{tr}[ \Psi ^*( \mathsf{X} ) ^{\dagger}\mathsf{Y} ] $ for two arbitrary matrices $\mathsf{X}$ and $\mathsf{Y}$. The optimum values associated with the primal and dual problems are respectively defined as
\begin{eqnarray}
    \alpha\coloneqq\sup\left\{\tr(\mathsf{A}\mathsf{X})|\mathsf{X}\geq 0, \Psi \left( \mathsf{X}  \right) =\mathsf{B}\right\},~~\beta\coloneqq\inf\left\{\tr(\mathsf{B}\mathsf{Y})|\Psi^* \left( \mathsf{Y}  \right) \ge\mathsf{A}, \mathsf{Y}=\mathsf{Y}^\dagger\right\}.
\end{eqnarray}
It is known that $\alpha\leq \beta$ and the equality is attained if Slater's conditions are met \cite{Watrous2018}. We will use the following Slater's theorem:
\begin{theorem}[Slater's theorem]
    If there exists a positive semidefinite operator $\mathsf{X}$ satisfying $\Psi \left( \mathsf{X}  \right) =\mathsf{B}$ and a Hermitian operator $\mathsf{Y}$ satisfying $\Psi^*(\mathsf{Y})>\mathsf{A}$, then $\alpha=\beta$ \cite{Watrous2018}.
\end{theorem}

With the above knowledge, we now prove Eq.~(\eqpsdp). Let us start by considering the following optimization problem:
\begin{eqnarray}\label{app_optimization_problem}
    \underset{h\in \mathbb{H} ^{\left( q \right)}}{\min}\,\,\mathrm{tr}\left[ \tilde{X}\Omega \left( h \right) \right],
\end{eqnarray}
with $\Omega(h)=4(H^*\Phi^*-i\Phi^* h)(H^*\Phi^*-i\Phi^* h)^\dagger$. We point out that this problem can be reformulated as the following SDP:
\begin{equation} \label{app_eq_min_origin}
    \begin{aligned}
        \underset{h,\Gamma}{\min}\quad & \mathrm{tr}\left( \tilde{X}\Gamma \right)
        \\
        \mathrm{s}.\mathrm{t}.\quad    & \left[ \begin{matrix}
                                                        \Gamma                                          & 2\left( H^*\Phi ^*-i\Phi ^*h \right) \\
                                                        2\left( H^*\Phi ^*-i\Phi ^*h \right) ^{\dagger} & \mathbb{I} _q                        \\
                                                    \end{matrix} \right] \ge 0,
    \end{aligned}
\end{equation}
where $\Gamma$ denotes a matrix of the same shape as $\tilde{X}$. This point can be verified by noting that the constraint in Eq.~(\ref{app_eq_min_origin}) can be equivalently formulated as $\Gamma\geq\Omega(h)$ according to the Schur complement condition for positive semidefiniteness \cite{Horn2017}.

We next write Eq.~(\ref{app_eq_min_origin}) in the standard form of a dual problem, i.e., \cref{app_sdp_dual}. We express $\mathsf{Y}$ in the form  $\mathsf{Y}=\left[ \begin{matrix}
            \mathsf{Y}_{11} & \mathsf{Y}_{12} \\
            \mathsf{Y}_{21} & \mathsf{Y}_{22} \\
        \end{matrix} \right]$, where $\mathsf{Y}_{11}$ and $\mathsf{Y}_{22}$ share the same shape with $\Gamma$ and $h$, respectively. We introduce the Hermitian-preserving map
\begin{equation}
    \Psi ^*\left( \mathsf{Y} \right) =\left[ \begin{matrix}
            \mathsf{Y}_{11}          & -2i\Phi ^*\mathsf{Y}_{22} \\
            2i\mathsf{Y}_{22}\Phi ^T & 0                         \\
        \end{matrix} \right],
\end{equation}
and define $\mathsf{A}, \mathsf{B}$ to be
\begin{equation} \label{app_eq_def_AB}
    \mathsf{A}=\left[ \begin{matrix}
            0                                      & -2H^*\Phi ^*   \\
            -2\left( H^*\Phi ^* \right) ^{\dagger} & -\mathbb{I} _q \\
        \end{matrix} \right] ,\quad \mathsf{B}=\left[ \begin{matrix}
            \tilde{X} & 0 \\
            0         & 0 \\
        \end{matrix} \right] .
\end{equation}
Simple algebra shows the equivalence between the SDP in Eq.~(\ref{app_eq_min_origin}) and the SDP in \cref{app_sdp_dual} with the above $\Psi^*$, $\mathsf{A}$, and $\mathsf{B}$.

We then write out the primal problem corresponding to the SDP in \cref{app_sdp_dual} with the $\Psi^*$, $\mathsf{A}$, and $\mathsf{B}$.
By the definition of the adjoint map,
\begin{equation}\label{app_sdp_primal_psi}
    \Psi \left( \mathsf{X} \right) =\left[ \begin{matrix}
            \mathsf{X}_{11} & 0                                                 \\
            0               & 2i\Phi ^T\mathsf{X}_{12}-2i\mathsf{X}_{21}\Phi ^* \\
        \end{matrix} \right] ,
\end{equation}
where $\mathsf{X}=\left[ \begin{matrix}
            \mathsf{X}_{11} & \mathsf{X}_{12} \\
            \mathsf{X}_{21} & \mathsf{X}_{22} \\
        \end{matrix} \right] $ is partitioned in the same way as $\mathsf{Y}$. Inserting Eqs.~(\ref{app_eq_def_AB}) and (\ref{app_sdp_primal_psi}) into \cref{app_sdp_primal}, we have that the primal problem is
\begin{equation}
    \begin{aligned}
        \underset{\mathsf{X}}{\max}\quad & -\mathrm{tr}\mathsf{X}_{22}-4\mathrm{Re}[\mathrm{tr}\left( H^*\Phi ^*\mathsf{X}_{21} \right)]
        \\
        \mathrm{s}.\mathrm{t}.\quad      & \left[ \begin{matrix}
                                                          \mathsf{X}_{11} & 0                                                 \\
                                                          0               & 2i\Phi ^T\mathsf{X}_{12}-2i\mathsf{X}_{21}\Phi ^* \\
                                                      \end{matrix} \right]=\left[ \begin{matrix}
                                                                                      \tilde{X} & 0 \\
                                                                                      0         & 0 \\
                                                                                  \end{matrix} \right] ,
        \\
                                         & \left[ \begin{matrix}
                                                          \mathsf{X}_{11} & \mathsf{X}_{12} \\
                                                          \mathsf{X}_{21} & \mathsf{X}_{22} \\
                                                      \end{matrix} \right]\ge 0,
    \end{aligned}
\end{equation}
which can be simplified to be
\begin{equation} \label{app_eq_beforesimplify}
    \begin{aligned}
        \underset{\mathsf{X}_{21},\mathsf{X}_{22}}{\max}\quad & -\mathrm{tr}\mathsf{X}_{22}-4\mathrm{Re}[\mathrm{tr}\left( H^*\Phi ^*\mathsf{X}_{21} \right)]
        \\
        \mathrm{s}.\mathrm{t}.\quad
                                                              & \left[ \begin{matrix}
                                                                               \tilde{X}       & \mathsf{X}_{21}^\dagger \\
                                                                               \mathsf{X}_{21} & \mathsf{X}_{22}         \\
                                                                           \end{matrix} \right]\ge 0,
        \\
                                                              & \mathsf{X}_{21}\Phi^* ~~\textrm{is Hermitian.}
    \end{aligned}
\end{equation}
Notably, the Slater's conditions are met. On one hand, the $\mathsf{X}$ with $\mathsf{X}_{11}=\tilde X$ and $\mathsf{X}_{12}=\mathsf{X}_{21}=0$ satisfies the condition $\Psi(\mathsf{X})=\mathsf{B}$. On the other hand, the condition $\Psi^*(\mathsf{Y})>\mathsf{A}$ can be met by the $\mathsf{Y}$ with $\mathsf{Y}_{11}=c \mathbb I_{IO}$ for a sufficiently large $c\in\mathbb R$.

Now, by the equivalence between the two problems in Eqs.~(\ref{app_optimization_problem}) and (\ref{app_eq_beforesimplify}), we can convert the formula in the main text
\begin{eqnarray}\label{app_eq_startplace}
    \mathcal{J} _{\max}^{\left( k \right)}=\underset{\tilde{X}\in \tilde{\mathbb{X}}^{\left( k \right)}}{\max}\,\,\underset{h\in \mathbb{H} ^{\left( q \right)}}{\min}\,\,\mathrm{tr}\left[ \tilde{X}\Omega \left( h \right) \right]
\end{eqnarray}
into the following SDP:
\begin{equation} \label{app_eq_beforesimplify_2}
    \begin{aligned}
        \underset{\tilde{X},\mathsf{X}_{21},\mathsf{X}_{22}}{\max}\quad & -\mathrm{tr}\mathsf{X}_{22}-4\mathrm{Re}[\mathrm{tr}\left( H^*\Phi ^*\mathsf{X}_{21} \right)]
        \\
        \mathrm{s}.\mathrm{t}.\quad                                     & \tilde{X}\ge 0,\quad \Lambda^{({k})}(\tilde{X}) =\tilde{X},\quad\mathrm{tr}\tilde{X}=d_O,
        \\&
        \left[ \begin{matrix}
                       \tilde{X}       & \mathsf{X}_{21}^\dagger \\
                       \mathsf{X}_{21} & \mathsf{X}_{22}         \\
                   \end{matrix} \right]\ge 0,
        \\
                                                                        & \mathsf{X}_{21}\Phi^* ~~\textrm{is Hermitian.},
    \end{aligned}
\end{equation}
which can be simplified to be\footnote{Note that the constraint $\tilde{X}\geq 0$ is redundant.}
\begin{equation} \label{app_eq_aftersimplify}
    \begin{aligned}
        \underset{\tilde{X},\mathsf{X}_{21},\mathsf{X}_{22}}{\max}\quad & -\mathrm{tr}\mathsf{X}_{22}-4\mathrm{Re}[\mathrm{tr}\left( H^*\Phi ^*\mathsf{X}_{21} \right)]
        \\
        \mathrm{s}.\mathrm{t}.\quad                                     & \Lambda^{({k})}(\tilde{X}) =\tilde{X},\quad\mathrm{tr}\tilde{X}=d_O,
        \\
                                                                        & \left[ \begin{matrix}
                                                                                         \tilde{X}       & \mathsf{X}_{21}^\dagger \\
                                                                                         \mathsf{X}_{21} & \mathsf{X}_{22}         \\
                                                                                     \end{matrix} \right]\ge 0,
        \\&\mathsf{X}_{21}\Phi^* ~\textrm{is Hermitian}.
    \end{aligned}
\end{equation}
Equation (\ref{app_eq_aftersimplify}) is just Eq.~(\eqpsdp) in the main text, where we have used $B$ and $C$ to represent  $\mathsf{X}_{21}$ and $\mathsf{X}_{22}$, respectively.

\section{Algorithm 1}\label{Algorithm1}

Here we present an algorithm for finding a reliable lower bound on $\mathcal{J}_{\max}^{(k)}$ for $k=i, ii, iv$. Note that the primal SDP can numerically find approximately optimal $\tilde{X}$, $B$, and $C$. However,
due to numerical errors, the $\tilde{X}$, $B$, and $C$ thus found, generally speaking, do not strictly satisfy the constraints in the primal SDP. This makes the result computed from the primal SDP not reliable. The working principle of our algorithm is to use symbolic calculations with MATHEMATICA to render the numerically found $\tilde{X}$, $B$, and $C$ to fulfill the constraints. The details can be found in \cref{alg_primal_lowerbound}.

\RestyleAlgo{ruled}
\begin{algorithm}[ht]
    \caption{Find a reliable lower bound on $\mathcal{J}^{(k)}_{\mathrm{max}}$ for $k=i, ii, iv$.}\label{alg_primal_lowerbound}
    \KwIn{Approximately optimal $\tilde{X}$, $B$, and $C$ obtained from the primal SDP in Eq.~(\eqpsdp)}
    \KwOut{Lower bound on $\mathcal{J}^{(k)}_{\mathrm{max}}$}
    $\tilde{X}^{(\mathrm{trun})},B^{(\mathrm{trun})}, C^{(\mathrm{trun})}\gets$ Truncate $\tilde X,B,C$\;

    $\tilde{X}^{\left( \mathrm{trun} \right)}\gets \frac{\tilde{X}^{\left( \mathrm{trun} \right)}+\tilde{X}^{\left( \mathrm{trun} \right) \dagger}}{2}$; $\quad C^{\left( \mathrm{trun} \right)}\gets \frac{C^{\left( \mathrm{trun} \right)}+C^{\left( \mathrm{trun} \right) \dagger}}{2}$\;

    $B^{\left( \mathrm{acc} \right)}\gets\frac{B^{\left( \mathrm{trun} \right)}\Phi ^*+\Phi ^TB^{\left( \mathrm{trun} \right) \dagger}}{2}\Phi ^{*-1}+B^{\left( \mathrm{trun} \right)}\left( \mathbb{I}_{IO} -\Phi ^*\Phi ^{*-1} \right) $\;

    $\tilde{X}^{(\mathrm{acc})}\gets \Lambda^{(k)}(\tilde X^{(\mathrm{trun})})$; $\quad \tilde{X}^{\left( \mathrm{acc} \right)}\gets \frac{\tilde{X}^{\left( \mathrm{acc} \right)}}{\mathrm{tr}\left[ \tilde{X}^{\left( \mathrm{acc} \right)} \right]}d_O$ \;

    Adjust $\epsilon$ from $0$ to $1$ such that $\left( 1-\epsilon \right) \left[ \begin{matrix}
                \tilde{X}^{\left( \mathrm{acc} \right)} & B^{\left( \mathrm{acc} \right) \dagger} \\
                B^{\left( \mathrm{acc} \right)}         & C^{(\mathrm{trun})}                     \\
            \end{matrix} \right] +\epsilon \frac{\mathbb{I}_{IO}}{d_{I}}\oplus \mathbb{I}_q \ge 0$\;

    $\tilde{X}^{\left( \mathrm{acc} \right)}\gets \left( 1-\epsilon \right) \tilde{X}^{\left( \mathrm{acc} \right)}+\epsilon \frac{\mathbb{I}_{IO}}{d_I}$; $\quad B^{\left( \mathrm{acc} \right)}\gets \left( 1-\epsilon \right) B^{\left( \mathrm{acc} \right)}$; $\quad C^{\left( \mathrm{acc} \right)}\gets \left( 1-\epsilon \right) C^{(\mathrm{trun})}+\epsilon \mathbb{I}_q $\;

    Output the lower bound $-\mathrm{tr}C^{\left( \mathrm{acc} \right)}-4\mathrm{Re}\left[ \mathrm{tr}\left(H^* \Phi ^*B^{\left( \mathrm{acc} \right)} \right) \right] $.
\end{algorithm}
To help a general reader readily see how \cref{alg_primal_lowerbound} works, we present some explanatory notes below:

Step 1 allows us to work with fractions and avoid numerical imprecision.

Step 2 renders $\tilde{X}^{(\mathrm{trun})}$ and $C^{(\mathrm{trun})}$ to fulfill the Hermiticity condition.

Step 3 renders $B^{(\mathrm{acc})}$ to fulfill the condition that $B^{(\mathrm{acc})}\Phi^*$ is Hermitian. Here $\Phi^{*-1}$ is the Moore-Penrose inverse of $\Phi^*$. If $\Phi$ has linearly independent columns, which is the case here, there is
$\Phi^{*-1}\Phi^*=\mathbb I_q$. Using this equality, one can verify that $B^{(\mathrm{acc})}\Phi^*$ is Hermitian.

Step 4 renders $\tilde{X}^{(\mathrm{acc})}$ to satisfy the conditions $\Lambda^{(k)}(\tilde{X}^{(\mathrm{acc})})=\tilde{X}^{(\mathrm{acc})}$ and $\tr\tilde{X}^{(\mathrm{acc})}=d_O$.

Steps 5 and 6 render $\tilde{X}^{(\mathrm{acc})}$, $B^{(\mathrm{acc})}$, and $C^{(\mathrm{acc})}$ to satisfy $\left[ \begin{matrix}
            \tilde{X}^{\left( \mathrm{acc} \right)} & B^{\left( \mathrm{acc} \right) \dagger} \\
            B^{\left( \mathrm{acc} \right)}         & C^{(\mathrm{acc})}                      \\
        \end{matrix} \right]\geq 0$.

\section{Proof of Eq.~(\eqdsdp)}

Here we prove Eq.~(\eqdsdp) in the main text. Note that the set $\tilde{\mathbb{X}}^{(k)}$ is convex and compact\footnote{We consider finite-dimensional Hilbert spaces. Therefore, the compactness of the set $\tilde{\mathbb{X}}^{(k)}$ follows from the fact that the set is closed and bounded.}, the set $\mathbb{H}^{(q)}$ is convex, and the objective function $\tr[\tilde{X}\Omega(h)]$ is convex on $\mathbb{H}^{(q)}$ and concave on $\tilde{\mathbb{X}}^{(k)}$. According to Sion's minimax theorem \cite{Komiya1988},  we can interchange the minimization and maximization in the formula (\eqJmaxminopt) without affecting the result. Therefore, we have
\begin{equation}\label{formula-interchange}
    \mathcal{J} _{\max}^{\left( k \right)}=\underset{h\in \mathbb{H} ^{\left( q \right)}}{\min}\,\,\underset{\tilde{X}\in \tilde{\mathbb{X}}^{\left( k \right)}}{\max}\,\,\mathrm{tr}\left[ \tilde{X}\Omega \left( h \right) \right] .
\end{equation}
Let us consider for now the following maximization problem:
\begin{equation} \label{app_eq_minmax_step1}
    \begin{aligned}
        \underset{\tilde{X}}{\max}\quad & \mathrm{tr}\left[ \tilde{X}\Omega \left( h \right) \right]
        \\
        \mathrm{s}.\mathrm{t}. \quad    & \tilde{X}\ge 0,\quad\Lambda ^{\left( k \right)}\left( \tilde{X} \right) =\tilde{X},\quad\mathrm{tr}\tilde{X}=d_O.
    \end{aligned}
\end{equation}
By introducing the Hermitian-preserving map
\begin{equation}
    \Psi \left( \mathsf{X} \right) =\left[ \begin{matrix}
            \mathsf{X}-\Lambda ^{\left( k \right)}\left( \mathsf{X} \right) & 0                     \\
            0                                                               & \mathrm{tr}\mathsf{X} \\
        \end{matrix} \right]
\end{equation}
and defining $\mathsf{A},\mathsf{B}$ to be the following two Hermitian matrices
\begin{equation}
    \mathsf{A}=
    \Omega \left( h \right),\quad\mathsf{B}=\left[ \begin{matrix}
            0 & 0   \\
            0 & d_O \\
        \end{matrix} \right] ,
\end{equation}
we can write \cref{app_eq_minmax_step1} in the form of \cref{app_sdp_primal}. Then, the dual problem of \cref{app_eq_minmax_step1} can be found by noting that
\begin{equation}
    \Psi ^*\left( \mathsf{Y} \right) =\mathsf{Y}_{22}\mathbb{I}_{IO} +\mathsf{Y}_{11}-\Lambda ^{\left( k \right)}\left( \mathsf{Y}_{11} \right),
\end{equation}
where $\mathsf{Y}=\left[ \begin{matrix}
            \mathsf{Y}_{11} & \mathsf{Y}_{12} \\
            \mathsf{Y}_{21} & \mathsf{Y}_{22} \\
        \end{matrix} \right]$ is partitioned in the same way as $\mathsf{B}$ and we have used $\Lambda^{(k)*}=\Lambda^{(k)}$. Explicitly, the dual problem of \cref{app_eq_minmax_step1} reads
\begin{equation} \label{app_eq_prove_slateropt}
    \begin{aligned}
        \underset{\mathsf{Y}}{\min}\quad & d_O\mathsf{Y}_{22}
        \\
        \mathrm{s}.\mathrm{t}.\quad      & \mathsf{Y}_{22}\mathbb{I}_{IO} +\left( id-\Lambda ^{\left( k \right)} \right) \left( \mathsf{Y}_{11} \right) \ge \Omega \left( h \right) ,
        \\
                                         & \mathsf{Y}^{\dagger}=\mathsf{Y}.
    \end{aligned}
\end{equation}
Notably, the Slater's conditions are met. Indeed, on the one hand, since $\Lambda^{(k)}$ is unital, that is, $\Lambda^{(k)}(\mathbb{I}_{IO})=\mathbb{I}_{IO}$, we have that $\tilde X=\mathbb{I}_{IO}/d_I$ satisfies all the constrains in \cref{app_eq_minmax_step1}. On the other hand, the matrix inequality in \cref{app_eq_prove_slateropt} can always be met strictly by adjusting the value of $\mathsf{Y}_{22}$.
Letting $\lambda=d_O \mathsf{Y}_{22}$ and $\tilde{Y}=(id-\Lambda^{(k)})(\mathsf{Y}_{11})$ and noting that $\mathsf{Y}_{12}$ and $\mathsf{Y}_{21}$ are dummy variables and therefore can be removed, we can rewrite Eq.~(\ref{app_eq_prove_slateropt}) in the following form:
\begin{equation}\label{proof-19-before}
    \begin{aligned}
        \underset{\lambda ,\tilde{Y}}{\min}\quad & \lambda
        \\
        \mathrm{s}.\mathrm{t}.\quad              & \lambda \frac{\mathbb{I}_{IO}}{d_O}+\tilde{Y}\ge \Omega \left( h \right) ,
        \\
                                                 & \Lambda ^{\left( k \right)}\left( \tilde{Y} \right) =0.
    \end{aligned}
\end{equation}
Here we have used the fact that $\Lambda^{(k)}$ is a projection and hence $\Lambda^{(k)}(\tilde Y)=0$. By the definition  $\Omega(h)=4(H^*\Phi^*-i\Phi^* h)(H^*\Phi^*-i\Phi^* h)^\dagger$ and the Schur complement condition for positive semidefiniteness \cite{Horn2017}, we can rewrite Eq.~(\ref{proof-19-before})
as
\begin{equation} \label{proof-19-after}
    \begin{aligned}
        \underset{\lambda,\tilde{Y}}{\min}\quad & \lambda
        \\
        \mathrm{s}.\mathrm{t}.\quad             & \Lambda^{(k)} \left( \tilde{Y} \right) =0,
        \\
                                                & \left[ \begin{matrix}
                                                                 \lambda \frac{\mathbb{I}_{IO}}{d_O}+\tilde{Y}   & 2\left( H^*\Phi ^*-i\Phi ^*h \right) \\
                                                                 2\left( H^*\Phi ^*-i\Phi ^*h \right) ^{\dagger} & \mathbb{I}_{q}                       \\
                                                             \end{matrix} \right] \ge 0.
    \end{aligned}
\end{equation}
Substituting Eq.~(\ref{proof-19-after}) into Eq.~(\ref{formula-interchange}), we arrive at Eq.~(\eqdsdp) in the main text.

\section{Algorithm 2}

\RestyleAlgo{ruled}
\begin{algorithm}[ht]
    \caption{Find a reliable upper bound on $\mathcal{J}^{(k)}_{\mathrm{max}}$ for $k=i, ii, iv$}\label{alg_dual_upperbound}
    \KwIn{Approximately optimal $\tilde Y$, $\lambda$, and $h$ obtained from the dual SDP in Eq.~(\eqdsdp)}
    \KwOut{Upper bound on $\mathcal{J}^{(k)}_{\mathrm{max}}$}

    $\tilde{Y}^{\left( \mathrm{trun} \right)},\lambda ^{\left( \mathrm{trun} \right)},h^{\left( \mathrm{trun} \right)}\gets$ truncate $\tilde Y,\lambda, h$\;

    $\tilde{Y}^{\left( \mathrm{trun} \right)}\gets \frac{\tilde{Y}^{\left( \mathrm{trun} \right)}+\tilde{Y}^{\left( \mathrm{trun} \right) \dagger}}{2};\quad h^{\left( \mathrm{acc} \right)}\gets \frac{h^{\left( \mathrm{trun} \right)}+h^{\left( \mathrm{trun} \right) \dagger}}{2}$\;

    $\tilde{Y}^{\left( \mathrm{acc} \right)}\gets ( id-\Lambda ^{( k )} ) ( \tilde{Y}^{\left( \mathrm{trun} \right)} )$\;

    $\lambda^{(\mathrm{acc})}\gets$ Gradually increase the value of $\lambda^{(\mathrm{trun})}$ until
    $
        \left[ \begin{matrix}
                \lambda^{(\mathrm{turn})} \frac{\mathbb{I}_{IO}}{d_O}+\tilde{Y}^{\left( \mathrm{acc} \right)} & 2\left( H^*\Phi ^*-i\Phi ^*h^{\left( \mathrm{acc} \right)} \right) \\
                2\left( H^*\Phi ^*-i\Phi ^*h^{\left( \mathrm{acc} \right)} \right) ^{\dagger}                 & \mathbb{I}_q                                                       \\
            \end{matrix} \right] \ge 0
    $
    is met\;

    Output the upper bound $\lambda^{(\mathrm{acc})}$.
\end{algorithm}

Here we present \cref{alg_dual_upperbound} for finding a reliable upper bound on $\mathcal{J}_{\mathrm{max}}^{(k)}$ for $k=i,ii,iv$. Here, steps 1 and 2 are the same as in \cref{alg_primal_lowerbound}; step 3 renders $\tilde Y^{(\mathrm{acc})}$ to satisfy the condition $\Lambda^{(k)}(\tilde Y^{(\mathrm{acc})})=0$; and step 4 renders $\lambda^{(\mathrm{acc})}$ to satisfy
\begin{equation}
    \left[ \begin{matrix}
            \lambda^{(\mathrm{acc})} \frac{\mathbb{I}_{IO}}{d_O}+\tilde{Y}^{\left( \mathrm{acc} \right)} & 2\left( H^*\Phi ^*-i\Phi ^*h^{\left( \mathrm{acc} \right)} \right) \\
            2\left( H^*\Phi ^*-i\Phi ^*h^{\left( \mathrm{acc} \right)} \right) ^{\dagger}                & \mathbb{I}_q                                                       \\
        \end{matrix} \right] \ge 0    .
\end{equation}

\section{Discussions on strategies of type $iii$} \label{app_sec_strategiesiii}

Here we discuss the strategies of type $iii$. To avoid cumbersome notations, we consider the $N=2$ case in what follows. We point out that all the results to be presented in this section can be extended to the general $N$ case straightforwardly.

\subsection{Characterization of $\mathbb{\tilde{X}}^{(iii)}$}

We clarify that the sequential strategies characterized by Eqs.~(\ref{chara-seq-set}) and (\ref{chara-Lambda-Seq}) are with the causal order $1\prec 2$. Here, $1\prec 2$ stands for the processes in which the first channel is queried before the second one. Hereafter, we denote by $\tilde X^{(1\prec2)}$ an element of the set $\mathbb{\tilde{X}}^{(ii)}$ with the causal order $1\prec 2$. That is, $\tilde X^{(1\prec2)}$ satisfies $\tilde{X}^{(1\prec2)}\ge 0$, $\tr \tilde{X}^{(1\prec2)}=d_O$, and
\begin{equation}
    \Lambda^{(1\prec2)}(\tilde{X}^{(1\prec2)})\coloneqq\prescript{}{O_2}{\tilde{X}^{(1\prec2)}}-\prescript{}{I_2O_2}{\tilde{X}^{(1\prec2)}}+\prescript{}{O_1I_2O_2}{\tilde{X}^{(1\prec2)}}=\tilde X^{(1\prec2)}.
\end{equation}
Here we have used $\Lambda^{(1\prec 2)}$ to represent the map $\Lambda^{(ii)}$ with the causal order $1\prec 2$. Notably, sequential strategies can also be associated with the causal order $2\prec 1$, which corresponds to the processes in which the first channel is queried after the second one. The associated operator, denoted by $\tilde X^{(2\prec1)}$ hereafter, satisfies $\tilde{X}^{(2\prec1)}\ge 0$, $\tr \tilde{X}^{(2\prec1)}=d_O$, and
\begin{equation}
    \Lambda^{(2\prec1)}(\tilde{X}^{(2\prec1)})\coloneqq\prescript{}{O_1}{\tilde{X}^{(2\prec1)}}-\prescript{}{I_1O_1}{\tilde{X}^{(2\prec1)}}+\prescript{}{O_2I_1O_1}{\tilde{X}^{(2\prec1)}}=\tilde X^{(2\prec1)}.
\end{equation}
The strategies of type $iii$ correspond to the processes in which
the two channels are probed in a superposition of the causal orders $1\prec 2$ and $2\prec 1$. Accordingly, $\tilde{\mathbb{X}}^{\left( iii \right)}$ can be characterized as (see Eq.~(18) in Ref.~\cite{araujoWitnessingCausalNonseparability2015} and Eq.~(7) in Ref.~\cite{Bavaresco2021})
\begin{equation} \label{chara-strategies-iii}
    \tilde{\mathbb{X}}^{\left( iii \right)}\coloneqq \left\{ \tilde{X}| \tilde{X}=p\tilde{X}^{(1\prec 2)}+\left( 1-p \right) \tilde{X}^{(2\prec 1)}, 0\le p\le 1 \right\}.
\end{equation}
That is, $\tilde{\mathbb{X}}^{\left( iii \right)}$ consists of all the  $\tilde{X}$ that are convex combinations of $\tilde X^{(1\prec 2)}$ and $\tilde X^{(2\prec1)}$. It is worth noting that $\tilde{\mathbb{X}}^{\left( iii \right)}$ is convex and compact.

\subsection{Primal SDP}\label{subsec-primal-sdp}
To derive the primal SDP for the strategies of type $iii$, we make use of the following formula [i.e., Eq.~(\eqJmaxminopt) in the main text]:
\begin{equation}\label{formula-strategies-iii-maxmin}
    \mathcal{J} _{\max}^{\left( iii \right)}=\underset{\tilde{X}\in \tilde{\mathbb{X}}^{\left( iii \right)}}{\max}\,\,\underset{h\in \mathbb{H} ^{\left( q \right)}}{\min}\,\,\mathrm{tr}\left[ \tilde{X}\Omega \left( h \right) \right] .
\end{equation}
By the same reasoning as in deriving \cref{app_eq_beforesimplify}, we have that the optimization problem $\underset{h\in \mathbb{H} ^{\left( q \right)}}{\min}\,\,\mathrm{tr}\left[ \tilde{X}\Omega \left( h \right) \right]$ can be reformulated as
\begin{equation}\label{sdp-strategies-iii-1}
    \begin{aligned}
        \underset{B,C}{\max}\quad   & -\mathrm{tr}C-4\mathrm{Re}\left[ \mathrm{tr}\left( H^*\Phi ^*B \right) \right]
        \\
        \mathrm{s}.\mathrm{t}.\quad & \left[ \begin{matrix}
                                                     \tilde{X} & B^{\dagger} \\
                                                     B         & C           \\
                                                 \end{matrix} \right] \ge 0,
        \\
                                    & B\Phi ^*\,\,\mathrm{is}\,\,\mathrm{Hermitian}.
    \end{aligned}
\end{equation}
Substituting Eqs.~(\ref{chara-strategies-iii}) and (\ref{sdp-strategies-iii-1}) into formula (\ref{formula-strategies-iii-maxmin}), we have that $\mathcal{J} _{\max}^{\left( iii \right)}$ can be computed via the following SDP:
\begin{equation}
    \begin{aligned}
        \underset{\tilde{X},B,C}{\max}\quad & -\mathrm{tr}C-4\mathrm{Re}\left[ \mathrm{tr}\left( H^*\Phi ^*B \right) \right]
        \\
        \mathrm{s}.\mathrm{t}.\quad         & \tilde{X}=p\tilde{X}^{(1\prec 2)}+\left( 1-p \right) \tilde{X}^{(2\prec 1)},
        \\
                                            & \tilde{X}^{(1\prec 2)}\ge 0,\quad\tilde{X}^{(2\prec 1)}\ge 0,
        \\
                                            & \Lambda ^{(1\prec 2)}\left( \tilde{X}^{(1\prec 2)} \right) =\tilde{X}^{(1\prec 2)},\quad \Lambda ^{(2\prec 1)}\left( \tilde{X}^{(2\prec 1)} \right) =\tilde{X}^{(2\prec 1)},
        \\
                                            & \mathrm{tr}\left( \tilde{X}^{(1\prec 2)} \right) =d_O,\quad\mathrm{tr}\left( \tilde{X}^{(2\prec 1)} \right) =d_O,
        \\
                                            & 0\le p\le 1,
        \\
                                            & \left[ \begin{matrix}
                                                             \tilde{X} & B^{\dagger} \\
                                                             B         & C           \\
                                                         \end{matrix} \right] \ge 0,
        \\
                                            & B\Phi ^*\,\,\mathrm{is}\,\,\mathrm{Hermitian}.
    \end{aligned}
\end{equation}
Further, introducing $\tilde{X}_1=p\tilde{X}^{(1\prec 2)}$ and $\tilde{X}_2=\left( 1-p \right) \tilde{X}^{(2\prec 1)}$ to avoid the products of variables, we can simplify the above SDP to be
\begin{equation} \label{app_eq_primalsdpiii}
    \begin{aligned}
        \underset{\tilde{X}_1,\tilde{X}_2,B,C}{\max}\quad & -\mathrm{tr}C-4\mathrm{Re}\left[ \mathrm{tr}\left( H^*\Phi ^*B \right) \right]
        \\
        \mathrm{s}.\mathrm{t}.\quad                       & \tilde{X}_1\ge 0,\quad\tilde{X}_2\ge 0,
        \\
                                                          & \Lambda ^{(1\prec 2)}\left( \tilde{X}_1 \right) =\tilde{X}_1,\quad\Lambda ^{(2\prec 1)}\left( \tilde{X}_2 \right) =\tilde{X}_2,
        \\
                                                          & \mathrm{tr}\left( \tilde{X}_1 \right) +\mathrm{tr}\left( \tilde{X}_2 \right) =d_O,
        \\
                                                          & \left[ \begin{matrix}
                                                                           \tilde{X}_1+\tilde{X}_2 & B^{\dagger} \\
                                                                           B                       & C           \\
                                                                       \end{matrix} \right] \ge 0,
        \\
                                                          & B\Phi ^*\,\,\mathrm{is}\,\,\mathrm{Hermitian}.
    \end{aligned}
\end{equation}
Here we have dropped the redundant constraint $0\leq p\leq 1$.
Equation (\ref{app_eq_primalsdpiii}) is the primal SDP for the strategies of type $iii$.

\subsection{Dual SDP}

Using Sion's minimax theorem \cite{Komiya1988},  we can interchange the minimization and maximization in Eq.~(\ref{formula-strategies-iii-maxmin}) and obtain
\begin{equation}\label{formula-strategies-iii-minmax}
    \mathcal{J} _{\max}^{\left( iii \right)}=\underset{h\in \mathbb{H} ^{\left( q \right)}}{\min}\,\,\underset{\tilde{X}\in \tilde{\mathbb{X}}^{\left( iii \right)}}{\max}\,\,\mathrm{tr}\left[ \tilde{X}\Omega \left( h \right) \right].
\end{equation}
From Eq.~(\ref{chara-strategies-iii}), it follows that the maximization part $\underset{\tilde{X}\in \tilde{\mathbb{X}}^{\left( iii \right)}}{\max}\,\,\mathrm{tr}\left[ \tilde{X}\Omega \left( h \right) \right]$ appearing in Eq.~(\ref{formula-strategies-iii-minmax}) can be formulated as
\begin{equation}\label{dual-sdp-strategies-iii-1}
    \begin{aligned}
        \underset{\tilde{X}}{\max}\quad & \mathrm{tr}\left[ \tilde{X}\Omega \left( h \right) \right]
        \\
        \mathrm{s}.\mathrm{t}.\quad     & \tilde{X}=p\tilde{X}^{(1\prec 2)}+\left( 1-p \right) \tilde{X}^{(2\prec 1)},
        \\
                                        & \tilde{X}^{(1\prec 2)}\ge 0,\quad \tilde{X}^{(2\prec 1)}\ge 0,
        \\
                                        & \Lambda ^{(1\prec 2)}\left( \tilde{X}^{(1\prec 2)} \right) =\tilde{X}^{(1\prec 2)}, \quad\Lambda ^{(2\prec 1)}\left( \tilde{X}^{(2\prec 1)} \right) =\tilde{X}^{(2\prec 1)},
        \\
                                        & \mathrm{tr}\left( \tilde{X}^{(1\prec 2)} \right) =d_O, \quad\mathrm{tr}\left( \tilde{X}^{(2\prec 1)} \right) =d_O,
        \\
                                        & 0\le p\le 1.
    \end{aligned}
\end{equation}
As done in Subsec.~\ref{subsec-primal-sdp}, we introduce $\tilde{X}_1=p\tilde{X}^{(1\prec 2)}$ and $\tilde{X}_2=(1-p)\tilde{X}^{(2\prec 1)}$, which allows us to simplify Eq.~(\ref{dual-sdp-strategies-iii-1}) as follows:
\begin{equation} \label{app_eq_readytobestandard}
    \begin{aligned}
        \underset{\tilde{X}_1,\tilde{X}_2}{\max}\quad & \mathrm{tr}\left[ \left( \tilde{X}_1+\tilde{X}_2 \right) \Omega \left( h \right) \right]
        \\
        \mathrm{s}.\mathrm{t}. \quad                  & \tilde{X}_1\ge 0,\quad \tilde{X}_2\ge 0,
        \\
                                                      & \Lambda ^{(1\prec 2)}\left( \tilde{X}_1 \right) =\tilde{X}_1,\quad \Lambda ^{(2\prec 1)}\left( \tilde{X}_2 \right) =\tilde{X}_2,
        \\
                                                      & \mathrm{tr}\left( \tilde{X}_1 \right) +\mathrm{tr}\left( \tilde{X}_2 \right) =d_O.
    \end{aligned}
\end{equation}
To proceed, we need to express \cref{app_eq_readytobestandard} in the standard form of a primal problem, i.e., \cref{app_sdp_primal}. To do this, we write $\mathsf{X}$ as
\begin{eqnarray}
    \left[ \begin{matrix}
            \mathsf{X}_{11} & \mathsf{X}_{12} \\
            \mathsf{X}_{21} & \mathsf{X}_{22} \\
        \end{matrix} \right],
\end{eqnarray}
where $\mathsf{X}_{11}$ and $\mathsf{X}_{22}$ are of the same shape as $\tilde{X}_1$ and $\tilde{X}_2$. We introduce the Hermitian-preserving map
\begin{equation}
    \Psi \left( \mathsf{X} \right) =\left[ \begin{matrix}
            \mathsf{X}_{11}-\Lambda ^{(1\prec 2)}\left( \mathsf{X}_{11} \right) & 0                                                                   & 0                                                                                    \\
            0                                                                   & \mathsf{X}_{22}-\Lambda ^{(2\prec 1)}\left( \mathsf{X}_{22} \right) & 0                                                                                    \\
            0                                                                   & 0                                                                   & \mathrm{tr}\left( \mathsf{X}_{11} \right) +\mathrm{tr}\left( \mathsf{X}_{22} \right) \\
        \end{matrix} \right]  ,
\end{equation}
and further define $\mathsf{A}$ and $\mathsf{B}$ to be the following two Hermitian matrices
\begin{equation}
    \mathsf{A}=\left[ \begin{matrix}
            \Omega \left( h \right) & 0                       \\
            0                       & \Omega \left( h \right) \\
        \end{matrix} \right] ,\quad\mathsf{B}=\left[ \begin{matrix}
            0 & 0 & 0   \\
            0 & 0 & 0   \\
            0 & 0 & d_O \\
        \end{matrix} \right].
\end{equation}
Simple algebra shows the equivalence between the SDP in Eq.~(\ref{app_eq_readytobestandard}) and the SDP in \cref{app_sdp_primal} with the above $\Psi$, $\mathsf{A}$, and $\mathsf{B}$. We then identify the corresponding dual problem. To this end, we write $\mathsf{Y}$ as
\begin{eqnarray}
    \left[ \begin{matrix}
            \mathsf{Y}_{11} & \mathsf{Y}_{12} & \mathsf{Y}_{13} \\
            \mathsf{Y}_{21} & \mathsf{Y}_{22} & \mathsf{Y}_{23} \\
            \mathsf{Y}_{31} & \mathsf{Y}_{32} & \mathsf{Y}_{33} \\
        \end{matrix} \right],
\end{eqnarray}
where $\mathsf{Y}_{11}$ and $\mathsf{Y}_{22}$ are of the same shape as $\mathsf{X}_{11}$ and $\mathsf{X}_{22}$, and $\mathsf{Y}_{33}\in\mathbb{R}$.
Noticing that
\begin{equation}
    \Psi ^*\left( \mathsf{Y} \right) =\left[ \begin{matrix}
            \mathsf{Y}_{33}\mathbb{I}_{IO} +\mathsf{Y}_{11}-\Lambda ^{(1\prec 2)
            }\left( \mathsf{Y}_{11} \right) & 0                                                                                                   \\
            0                               & \mathsf{Y}_{33}\mathbb{I}_{IO} +\mathsf{Y}_{22}-\Lambda ^{(2\prec 1)}\left( \mathsf{Y}_{22} \right) \\
        \end{matrix} \right] ,
\end{equation}
we have that the dual problem reads
\begin{equation}\label{strategies-iii-dual-sdp-1}
    \begin{aligned}
        \underset{\mathsf{Y}_{11},\mathsf{Y}_{22},\mathsf{Y}_{33}}{\min}\quad & \mathsf{Y}_{33}d_O
        \\
        \mathrm{s}.\mathrm{t}. \quad                                          & \mathsf{Y}_{33}\mathbb{I}_{IO} +\left( id-\Lambda ^{(1\prec 2)} \right) \left( \mathsf{Y}_{11} \right) \ge \Omega \left( h \right) ,
        \\
                                                                              & \mathsf{Y}_{33}\mathbb{I}_{IO} +\left( id-\Lambda ^{(2\prec 1)} \right) \left( \mathsf{Y}_{22} \right) \ge \Omega \left( h \right) .
    \end{aligned}
\end{equation}
Here we have removed the dummy variables $\mathsf{Y}_{12}$, $\mathsf{Y}_{13}$, $\mathsf{Y}_{21}$, $\mathsf{Y}_{23}$, $\mathsf{Y}_{31}$, and $\mathsf{Y}_{32}$. Further, rewriting Eq.~(\ref{strategies-iii-dual-sdp-1}) by
letting
\begin{eqnarray}
    \lambda=\mathsf{Y}_{33}d_O,~~\tilde{Y}_1=(id-\Lambda ^{(1\prec 2)})(\mathsf{Y}_{11}),~~\tilde{Y}_2=(id-\Lambda ^{(2\prec 1)})(\mathsf{Y}_{22}),
\end{eqnarray}
we have
\begin{equation}\label{sdp-strategies-iii-dual-2}
    \begin{aligned}
        \underset{\lambda ,\tilde{Y}_1,\tilde{Y}_2}{\min}\quad & \lambda
        \\
        \mathrm{s}.\mathrm{t}. \quad
                                                               & \lambda\frac{\mathbb{I}_{IO}}{d_O}+\tilde{Y}_1\ge \Omega \left( h \right) ,\quad\lambda\frac{\mathbb{I}_{IO}}{d_O}+\tilde{Y}_2\ge \Omega \left( h \right),
        \\
                                                               & \Lambda ^{(1\prec 2)}\left( \tilde{Y}_1 \right) =0, \quad\Lambda ^{(2\prec 1)}\left( \tilde{Y}_2 \right) =0.
    \end{aligned}
\end{equation}
Here we have used the fact that $\Lambda ^{(1\prec 2)}$ and $\Lambda ^{(2\prec 1)}$ are projections and therefore $\Lambda ^{(1\prec 2)}\left( \tilde{Y}_1 \right) =0$ and $\quad\Lambda ^{(2\prec 1)}\left( \tilde{Y}_2 \right) =0$. Notably, the Slater's conditions are met. Indeed, on the one hand, as $\Lambda^{(1\prec 2)}$ and $\Lambda^{(2\prec 1)}$ are unital, that is, $\Lambda^{(1\prec 2)}(\mathbb{I}_{IO})=\mathbb{I}_{IO}$ and $\Lambda^{(2\prec 1)}(\mathbb{I}_{IO})=\mathbb{I}_{IO}$, we have that the $\mathsf{X}$ with $\mathsf{X}_{11}=\mathsf{X}_{22}=\mathbb{I}_{IO}/2d_I$ and $\mathsf{X}_{12}=\mathsf{X}_{21}=0$  satisfies the constrains $\mathsf{X}\geq 0$ and $\Psi(\mathsf{X})=\mathsf{B}$. On the other hand, the constraint $\Psi^*(\mathsf{Y})\geq \mathsf{A}$ can be met strictly by adjusting the value of $\mathsf{Y}_{33}$. Now, we see that the maximization part ${\max}_{\tilde{X}\in \tilde{\mathbb{X}}^{\left( iii \right)}}\,\,\mathrm{tr}\left[ \tilde{X}\Omega \left( h \right) \right]$ appearing in formula (\ref{formula-strategies-iii-minmax}) can be computed via the SDP in Eq.~(\ref{sdp-strategies-iii-dual-2}). Finally, inserting Eq.~(\ref{sdp-strategies-iii-dual-2}) into formula (\ref{formula-strategies-iii-minmax}), we obtain that $\mathcal{J}_{\max}^{(iii)}$ can be computed as
\begin{equation} \label{app_eq_dualsdpiii}
    \begin{aligned}
        \underset{h,\lambda ,\tilde{Y}_1,\tilde{Y}_2}{\min}\quad & \lambda
        \\
        \mathrm{s}.\mathrm{t}.\quad                              & \Lambda ^{(1\prec 2)}\left( \tilde{Y}_1 \right) =0,\quad \Lambda ^{(2\prec 1)}\left( \tilde{Y}_2 \right) =0,
        \\
                                                                 & \left[ \begin{matrix}
                                                                                  \lambda\frac{\mathbb{I}_{IO}}{d_O}+\tilde{Y}_1  & 2\left( H^*\Phi ^*-i\Phi ^*h \right) \\
                                                                                  2\left( H^*\Phi ^*-i\Phi ^*h \right) ^{\dagger} & \mathbb{I}_q                         \\
                                                                              \end{matrix} \right] \ge 0,
        \\
                                                                 & \left[ \begin{matrix}
                                                                                  \lambda\frac{\mathbb{I}_{IO}}{d_O}+\tilde{Y}_2  & 2\left( H^*\Phi ^*-i\Phi ^*h \right) \\
                                                                                  2\left( H^*\Phi ^*-i\Phi ^*h \right) ^{\dagger} & \mathbb{I}_q                         \\
                                                                              \end{matrix} \right] \ge 0,
    \end{aligned}
\end{equation}
where we have used the Schur complement condition for positive semidefiniteness \cite{Horn2017}. Equation (\ref{app_eq_dualsdpiii}) is the dual SDP for strategies of type $iii$.

\subsection{Two algorithms}
Here we present \cref{alg_primal_lowerbound__striii} and \cref{alg_dual_upperbound_striii} for obtaining reliable lower and upper bounds on $\mathcal{J}_{\mathrm{max}}^{(iii)}$, respectively.

\RestyleAlgo{ruled}
\begin{algorithm}[ht]
    \caption{Find a reliable lower bound on $\mathcal{J}^{(iii)}_{\mathrm{max}}$}\label{alg_primal_lowerbound__striii}
    \KwIn{Approximately optimal $\tilde X_1$, $\tilde X_2$, $B$, and $C$ obtained from the primal SDP in Eq.~(\ref{app_eq_primalsdpiii})}
    \KwOut{Lower bound on $\mathcal{J}^{(iii)}_{\mathrm{max}}$}

    $\tilde X_1^{(\mathrm{trun})},\tilde X_2^{(\mathrm{trun})},B^{(\mathrm{trun})},C^{(\mathrm{trun})}\gets$ truncate $\tilde X_1,\tilde X_2,B,C$\;

    $\tilde{X}_{1}^{\left( \mathrm{trun} \right)}\gets \frac{\tilde{X}_{1}^{\left( \mathrm{trun} \right)}+\tilde{X}_{1}^{\left( \mathrm{trun} \right) \dagger}}{2};\quad\tilde{X}_{2}^{\left( \mathrm{trun} \right)}\gets \frac{\tilde{X}_{2}^{\left( \mathrm{trun} \right)}+\tilde{X}_{2}^{\left( \mathrm{trun} \right) \dagger}}{2};\quad C^{\left( \mathrm{trun} \right)}\gets \frac{C^{\left( \mathrm{trun} \right)}+C^{\left( \mathrm{trun} \right) \dagger}}{2}$\;

    $B^{\left( \mathrm{acc} \right)}\gets \frac{B^{\left( \mathrm{trun} \right)}\Phi ^*+\Phi ^TB^{\left( \mathrm{trun} \right) \dagger}}{2}\Phi ^{*-1}+B^{\left( \mathrm{trun} \right)}\left( \mathbb{I}_{IO} -\Phi ^*\Phi ^{*-1} \right) $\;

    $\tilde{X}_{1}^{\left( \mathrm{acc} \right)}\gets \Lambda ^{(1\prec 2)}( \tilde{X}_{1}^{\left( \mathrm{trun} \right)} ) ;\quad\tilde{X}_{2}^{\left( \mathrm{acc} \right)}\gets \Lambda ^{(2\prec 1)}( \tilde{X}_{2}^{\left( \mathrm{trun} \right)} ) $\;

    $\tilde{X}_{1}^{\left( \mathrm{acc} \right)}\gets \frac{\tilde{X}_{1}^{\left( \mathrm{acc} \right)}d_O}{\mathrm{tr}\tilde{X}_{1}^{\left( \mathrm{acc} \right)}+\mathrm{tr}\tilde{X}_{2}^{\left( \mathrm{acc} \right)}};\quad\tilde{X}_{2}^{\left( \mathrm{acc} \right)}\gets \frac{\tilde{X}_{2}^{\left( \mathrm{acc} \right)}d_O}{\mathrm{tr}\tilde{X}_{1}^{\left( \mathrm{acc} \right)}+\mathrm{tr}\tilde{X}_{2}^{\left( \mathrm{acc} \right)}}$\;

    Adjust $\epsilon_1,\epsilon_2$ from $0$ to $1$ such that $\left( 1-\epsilon _1 \right) \tilde{X}_{1}^{\left( \mathrm{acc} \right)}+\epsilon _1\frac{\mathbb{I}_{IO}}{d_I} \ge 0$ and $\left( 1-\epsilon _2 \right) \tilde{X}_{2}^{\left( \mathrm{acc} \right)}+\epsilon _2\frac{\mathbb{I}_{IO}}{d_I} \ge 0$
    \;

    $\tilde{X}_{1}^{\left( \mathrm{acc} \right)}\gets \left( 1-\epsilon _1 \right) \tilde{X}_{1}^{\left( \mathrm{acc} \right)}+\epsilon _1\frac{\mathbb{I}_{IO}}{d_I};\quad\tilde{X}_{2}^{\left( \mathrm{acc} \right)}\gets \left( 1-\epsilon _2 \right) \tilde{X}_{2}^{\left( \mathrm{acc} \right)}+\epsilon _2\frac{\mathbb{I}_{IO}}{d_I} $\;

    Adjust $\epsilon$ such that $\left( 1-\epsilon \right) \left[ \begin{matrix}
                \tilde{X}_{1}^{\left( \mathrm{acc} \right)}+\tilde{X}_{2}^{\left( \mathrm{acc} \right)} & B^{\left( \mathrm{acc} \right) \dagger} \\
                B^{\left( \mathrm{acc} \right)}                                                         & C^{\left( \mathrm{trun} \right)}        \\
            \end{matrix} \right] +\epsilon \frac{\mathbb{I}_{IO}}{d_I}\oplus \mathbb{I}_q \ge 0$\;

    $B^{( \mathrm{acc} )}\gets\left( 1-\epsilon \right) B^{\left( \mathrm{acc} \right)};\quad C^{\left( \mathrm{acc} \right)}\gets ( 1-\epsilon ) C^{\left( \mathrm{trun} \right)}+\epsilon \mathbb{I}_q$\;

    %$\tilde{X}_{1}^{\left( \mathrm{acc} \right)}=\left( 1-\epsilon \right) \tilde{X}_{1}^{\left( \mathrm{acc} \right)}+\epsilon\frac{ \mathrm{tr}\tilde{X}_{1}^{\left( \mathrm{acc} \right)}}{\mathrm{tr}\tilde{X}_{1}^{\left( \mathrm{acc} \right)}+\mathrm{tr}\tilde{X}_{2}^{\left( \mathrm{acc} \right)}}\frac{\mathbb{I}_{IO}}{d_I};\quad \tilde{X}_{2}^{\left( \mathrm{acc} \right)}=\left( 1-\epsilon \right) \tilde{X}_{2}^{\left( \mathrm{acc} \right)}+\epsilon\frac{ \mathrm{tr}\tilde{X}_{2}^{\left( \mathrm{acc} \right)}}{\mathrm{tr}\tilde{X}_{1}^{\left( \mathrm{acc} \right)}+\mathrm{tr}\tilde{X}_{2}^{\left( \mathrm{acc} \right)}}\frac{\mathbb{I}_{IO}}{d_I}$\;

    Output the lower bound $-\mathrm{tr}C^{(\mathrm{acc})}-4\mathrm{Re}\left[ \mathrm{tr}\left( H^*\Phi ^*B^{(\mathrm{acc})} \right) \right] $.
\end{algorithm}

To help a general reader readily see how \cref{alg_primal_lowerbound__striii} works, we present some explanatory notes below:

Step 1 allows us to work with fractions and avoid numerical imprecision.

Step 2 renders $\tilde{X}_1^{(\mathrm{trun})}$, $\tilde{X}_2^{(\mathrm{trun})}$, and $C^{(\mathrm{trun})}$ to fulfill the Hermiticity condition.

Step 3 renders $B^{(\mathrm{acc})}$ to fulfill the condition that $B^{(\mathrm{acc})}\Phi^*$ is Hermitian. As in \cref{alg_primal_lowerbound}, $\Phi^{*-1}$ is the Moore-Penrose inverse of $\Phi^*$.

Steps 4 and 5 render $\tilde{X}_1^{(\mathrm{acc})}$ and $\tilde{X}_2^{(\mathrm{acc})}$ to satisfy the conditions $\Lambda^{(1\prec 2)}(\tilde{X}_1^{(\mathrm{acc})})=\tilde{X}_1^{(\mathrm{acc})}$, $\Lambda^{(2\prec 1)}(\tilde{X}_2^{(\mathrm{acc})})=\tilde{X}_2^{(\mathrm{acc})}$, and $\tr[\tilde{X}_1^{(\mathrm{acc})}+\tilde{X}_2^{(\mathrm{acc})}]=d_O$.

Steps 6 and 7 render $\tilde{X}_1^{(\mathrm{acc})}$ and $\tilde{X}_2^{(\mathrm{acc})}$ to be positive semidefinite.

\RestyleAlgo{ruled}
\begin{algorithm}[ht]
    \caption{Find a reliable upper bound on $\mathcal{J}^{(iii)}_{\mathrm{max}}$}\label{alg_dual_upperbound_striii}
    \KwIn{Approximately optimal $\tilde Y_1$, $\tilde Y_2$, $h$, and $\lambda$ obtained from the dual SDP in Eq.~(\ref{app_eq_dualsdpiii})}
    \KwOut{Upper bound on $\mathcal{J}^{(iii)}_{\mathrm{max}}$}

    $\tilde{Y}_{1}^{\left( \mathrm{trun} \right)},\tilde{Y}_{2}^{\left( \mathrm{trun} \right)},\lambda ^{\left( \mathrm{trun} \right)},h^{\left( \mathrm{trun} \right)}\gets$ truncate $\tilde Y_1,\tilde Y_2,\lambda,h$\;

    $\tilde{Y}_{1}^{\left( \mathrm{trun} \right)}\gets \frac{\tilde{Y}_{1}^{\left( \mathrm{trun} \right)}+\tilde{Y}_{1}^{\left( \mathrm{trun} \right) \dagger}}{2};\quad \tilde{Y}_{2}^{\left( \mathrm{trun} \right)}\gets \frac{\tilde{Y}_{2}^{\left( \mathrm{trun} \right)}+\tilde{Y}_{2}^{\left( \mathrm{trun} \right) \dagger}}{2};\quad h^{\left( \mathrm{trun} \right)}=\frac{h^{\left( \mathrm{trun} \right)}+h^{\left( \mathrm{trun} \right) \dagger}}{2}$\;

    $\tilde{Y}_{1}^{\left( \mathrm{acc} \right)}\gets ( id-\Lambda ^{(1\prec 2)} ) ( \tilde{Y}_{1}^{\left( \mathrm{trun} \right)});\quad \tilde{Y}_{2}^{\left( \mathrm{acc} \right)}\gets( id-\Lambda ^{(2\prec 1)} ) ( \tilde{Y}_{2}^{\left( \mathrm{trun} \right)} )$\;

    $\lambda^{(\mathrm{acc})}\gets$ Gradually increase the value of $\lambda^{(\mathrm{trun})}$ until
    \begin{equation*}
        \left[ \begin{matrix}
                \lambda ^{\left( \mathrm{trun} \right)}\frac{\mathbb{I}_{IO}}{d_O}+\tilde{Y}_{1}^{\left( \mathrm{acc} \right)} & 2\left( H^*\Phi ^*-i\Phi ^*h^{(\mathrm{acc})} \right) \\
                2\left( H^*\Phi ^*-i\Phi ^*h^{(\mathrm{acc})} \right) ^{\dagger}                                               & \mathbb{I}_q                                          \\
            \end{matrix} \right] \ge 0,\quad \left[ \begin{matrix}
                \lambda ^{\left( \mathrm{trun} \right)}\frac{\mathbb{I}_{IO}}{d_O}+\tilde{Y}_{2}^{\left( \mathrm{acc} \right)} & 2\left( H^*\Phi ^*-i\Phi ^*h^{(\mathrm{acc})} \right) \\
                2\left( H^*\Phi ^*-i\Phi ^*h^{(\mathrm{acc})} \right) ^{\dagger}                                               & \mathbb{I}_q                                          \\
            \end{matrix} \right] \ge 0
    \end{equation*}
    are met\;

    Output the upper bound $\lambda^{(\mathrm{acc})}$.
\end{algorithm}

\section{Method for randomly generating channels}

To study the hierarchy phenomenon reported in Theorem $\thH$, we have randomly generated 1000 channels in the main text.
Here we elaborate on the technical details of the method employed to generate these channels. All of the channels generated are of the form $\mathcal{E}_\theta=\mathcal{E}\circ \mathcal{U}_\theta$, where $\mathcal{U}_\theta$ is the unitary channel defined in the main text and $\mathcal{E}$ denotes a channel randomly generated by our method. Specifically, to randomly generate $\mathcal{E}$, we resort to the Stinespring dilation of $\mathcal{E}$ \cite{Watrous2018}, that is, $\mathcal{E}(\rho)=\mathrm{tr}_{\mathrm{aux}}\left( U\rho \otimes |0\rangle \langle 0|U^{\dagger} \right) $, where $|0\rangle \langle 0|$ is the initial state of an auxiliary system and $U$ denotes the unitary evolution operator of the system in question and the auxiliary system. Through generating unitary operators $U$ randomly according to the Haar measure \cite{qetlab}, we can obtain randomly generated channels $\mathcal{E}$.

\section{Examining different priors}

Here we examine the ultimate performance of global
estimation strategies ($i$)-($iv$) with different prior knowledge $p(\theta)$. To numerically solve SDPs in this work, we use CVX, a package for specifying and solving convex programs \cite{Grant2014,Grant2008}.

\begin{figure}
    \centering
    \includegraphics[width=\linewidth]{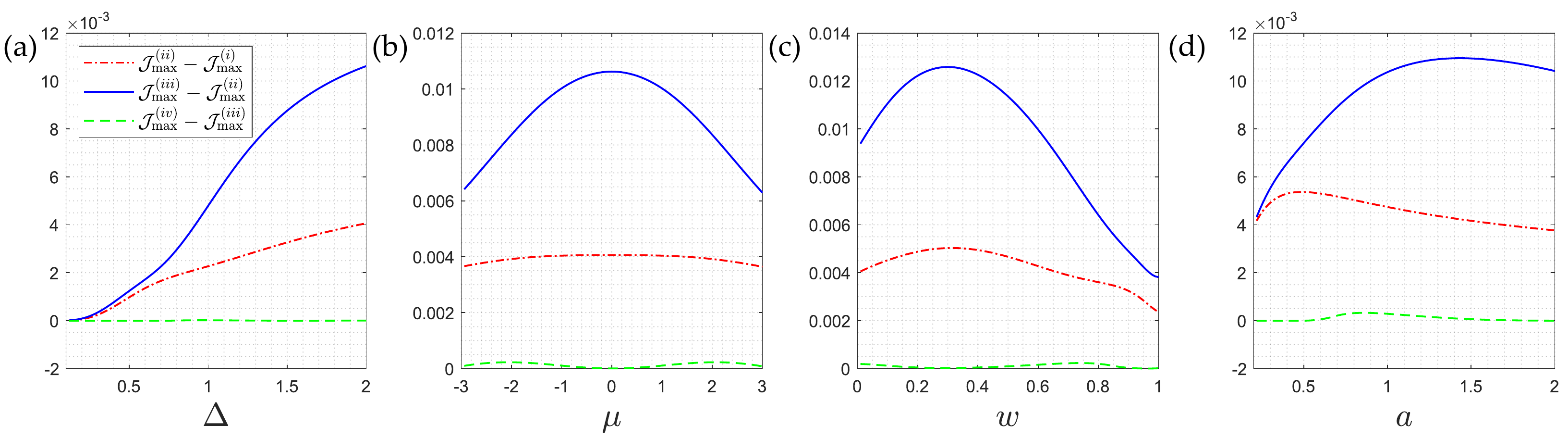}
    \caption{Numerical results displaying the influence of different priors $p(\theta)$ on the gaps between the maximal information $\mathcal{J}_{\max}^{(i)}$,  $\mathcal{J}_{\max}^{(ii)}$, $\mathcal{J}_{\max}^{(iii)}$, and $\mathcal{J}_{\max}^{(iv)}$. We consider four priors: (a) the Gaussian distribution with fixed mean value $\mu=0$ and varied standard deviation $\Delta$, (b) the Gaussian distribution with varied mean value $\mu$ and fixed standard deviation $\Delta=2$, (c) the Gaussian mixture distributions with varied weight $w$, and (d) the Beta distributions with varied $a$. We plot the gaps $\mathcal{J}_{\max}^{(k+1)}-\mathcal{J}_{\max}^{(k)}$ as a function of (a) $\Delta$, (b) $\mu$, (c) $w$, and (d) $a$. The red, blue, and green curves correspond to the gaps $\mathcal{J}_{\max}^{(ii)}-\mathcal{J}_{\max}^{(i)}$, $\mathcal{J}_{\max}^{(iii)}-\mathcal{J}_{\max}^{(ii)}$, and $\mathcal{J}_{\max}^{(iv)}-\mathcal{J}_{\max}^{(iii)}$, respectively. The channel examined here is $\mathcal{E}_\theta=\mathcal{E}^{(\mathrm{AD})}\circ \mathcal{E}^{(\mathrm{BF})}\circ \mathcal{U}_\theta$ which is the same as that considered in the main text.}
    \label{fig_stricthierarchy_diffpd}
\end{figure}

Figure \ref{fig_stricthierarchy_diffpd} displays the influence of different priors $p(\theta)$ on the gaps between $\mathcal{J}_{\max}^{(i)}$,  $\mathcal{J}_{\max}^{(ii)}$, $\mathcal{J}_{\max}^{(iii)}$, and $\mathcal{J}_{\max}^{(iv)}$. The channel under consideration is $\mathcal{E}_\theta=\mathcal{E}^{(\mathrm{AD})}\circ \mathcal{E}^{(\mathrm{BF})}\circ \mathcal{U}_\theta$ which is the same as that considered in the main text. In Fig.~\ref{fig_stricthierarchy_diffpd}(a), we examine the Gaussian distribution with fixed mean value $\mu=0$ and varied standard deviation $\Delta$. That is, \begin{eqnarray}
    p(\theta)=\frac{1}{\sqrt{2\pi\Delta^2}}e^{-\frac{(\theta-\mu)^2}{2\Delta^2}},
\end{eqnarray}
and we set $\mu=0$ and examine the gap $\mathcal{J}_{\max}^{(k+1)}-\mathcal{J}_{\max}^{(k)}$ as a function of $\Delta$. In Fig.~\ref{fig_stricthierarchy_diffpd}(a) as well as in other subfigures, the red, blue, and green curves correspond to the three gaps $\mathcal{J}_{\max}^{(ii)}-\mathcal{J}_{\max}^{(i)}$, $\mathcal{J}_{\max}^{(iii)}-\mathcal{J}_{\max}^{(ii)}$, and $\mathcal{J}_{\max}^{(iv)}-\mathcal{J}_{\max}^{(iii)}$, respectively.

As can be seen from Fig.~\ref{fig_stricthierarchy_diffpd}(a), the three gaps coincide in the limit of $\Delta\rightarrow 0$. Note that the smaller the value of $\Delta$ is, the more the prior knowledge about $\theta$ we have. The coincidence means that the four types of global estimation strategies $(i)$-$(iv)$ perform equally well when we have substantial prior knowledge about $\theta$. However, in the course of increasing $\Delta$, the two gaps $\mathcal{J}_{\max}^{(ii)}-\mathcal{J}_{\max}^{(i)}$ and $\mathcal{J}_{\max}^{(iii)}-\mathcal{J}_{\max}^{(ii)}$ increase monotonically, whereas the gap $\mathcal{J}_{\max}^{(iv)}-\mathcal{J}_{\max}^{(iii)}$ remains unchanged.

Figure \ref{fig_stricthierarchy_diffpd}(b) examines the Gaussian distribution with varied mean value $\mu$ and fixed standard deviation $\Delta=2$. As can be seen from Fig.~\ref{fig_stricthierarchy_diffpd}(b), the two gaps $\mathcal{J}_{\max}^{(ii)}-\mathcal{J}_{\max}^{(i)}$ and $\mathcal{J}_{\max}^{(iv)}-\mathcal{J}_{\max}^{(iii)}$ do not change much in the course of varying $\mu$. By contrast, the gap $\mathcal{J}_{\max}^{(iii)}-\mathcal{J}_{\max}^{(ii)}$ first increases as $\mu$ increases from $-3$ to $0$ and then decreases when $\mu$ keeps increasing.

Figure \ref{fig_stricthierarchy_diffpd}(c) examines the Gaussian mixture distributions with varied weight $w$. Here a Gaussian mixture distribution \cite{Reynolds2015} is defined as
\begin{equation}
    f\left( \theta \right) =\sum_{i}{w_if\left( \theta |\mu _i,\Delta _i \right)},
\end{equation}
where $f(\theta|\mu_i,\Delta_i)=\frac{1}{\sqrt{2\pi \Delta _{i}^{2}}}\exp \left[ -\left( \theta -\mu _i \right) ^2/\left( 2\Delta _{i}^{2} \right) \right]$ is the Gaussian distribution. In Fig.~\ref{fig_stricthierarchy_diffpd}(c), we set the prior $p(\theta)$ to be
\begin{equation}
    p(\theta)\propto wf(\theta|\mu_1,\Delta_1)+(1-w)f(\theta|\mu_2,\Delta_2)
\end{equation}
with $\Delta_1=1$, $\Delta_2=2$, $\mu_1=-\pi/2$, and $\mu_2=\pi/2$. In the course of varying $w$, the two gaps $\mathcal{J}_{\max}^{(iii)}-\mathcal{J}_{\max}^{(ii)}$ and $\mathcal{J}_{\max}^{(iv)}-\mathcal{J}_{\max}^{(iii)}$ behave similarly as in Fig.~\ref{fig_stricthierarchy_diffpd}(b). However, the gap $\mathcal{J}_{\max}^{(ii)}-\mathcal{J}_{\max}^{(i)}$ slowly changes, as can be see from Fig.~\ref{fig_stricthierarchy_diffpd}(c).

Figure \ref{fig_stricthierarchy_diffpd}(d) examines the Beta distribution with varied $a$. That is, we set $p(\theta)$ to be the Beta distribution defined as
\begin{equation}
    p\left( \theta \right) \propto \left( \frac{\theta +\pi}{2\pi} \right) ^{a-1}\left( 1-\frac{\theta +\pi}{2\pi} \right) ^{b-1},
\end{equation}
where $\theta\in [-\pi,\pi)$, $b=2$, and $a$ is varied from $0.2$ to $2$. Overall, the three gaps behave similarly as in Fig.~\ref{fig_stricthierarchy_diffpd}(a), but with the difference that the two gaps $\mathcal{J}_{\max}^{(ii)}-\mathcal{J}_{\max}^{(i)}$ and $\mathcal{J}_{\max}^{(iii)}-\mathcal{J}_{\max}^{(ii)}$ first increase and then decrease in course of increasing $a$.

\begin{figure}
    \centering
    \includegraphics[width=0.5\linewidth]{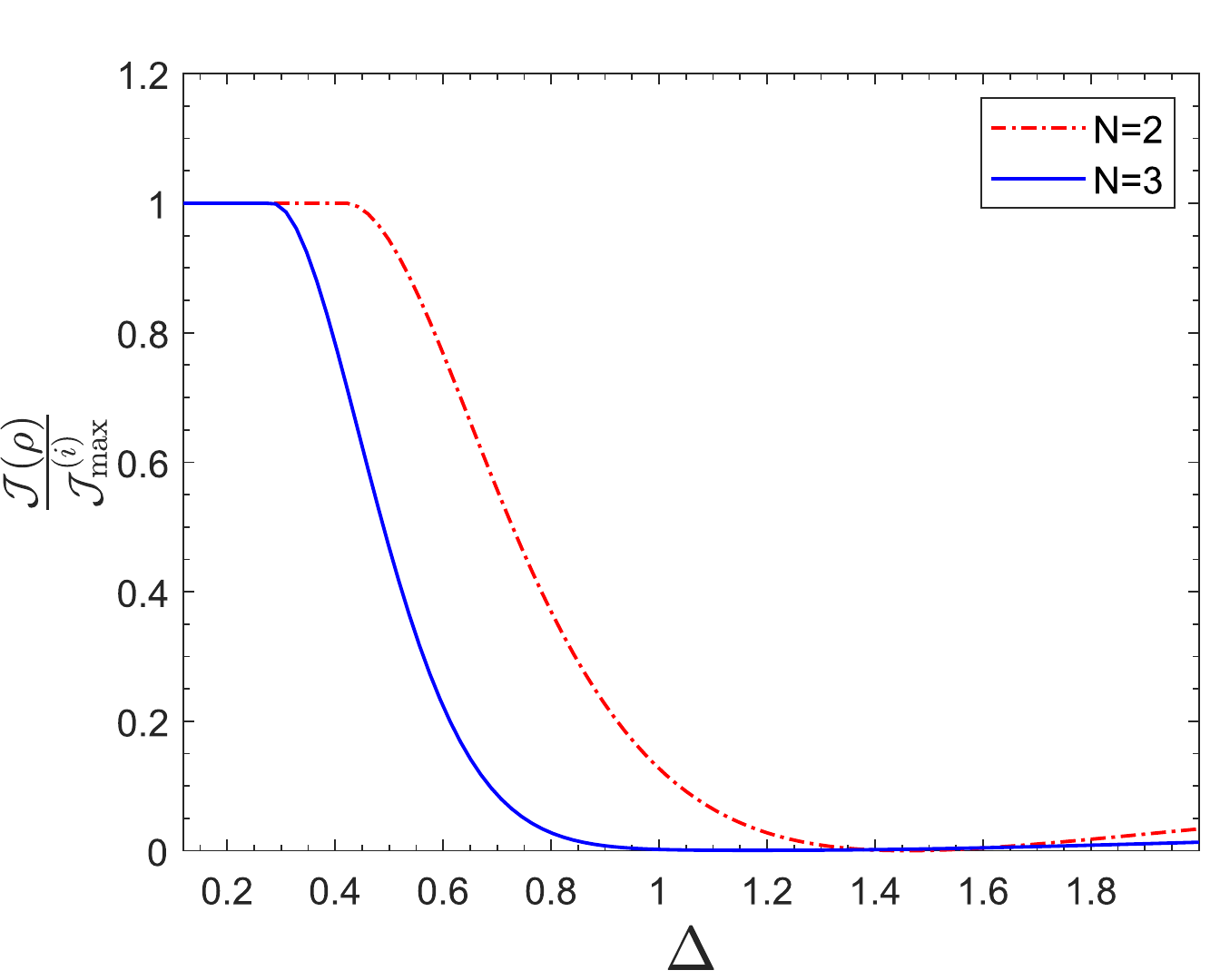}
    \caption{Numerical results displaying the influence of different priors $p(\theta)$ on the information gained in the global parallel strategy whose input state $\rho$ is chosen to be the GHZ state $\left( |0\rangle ^{\otimes N}+|1\rangle ^{\otimes N} \right) /\sqrt{2}$. Here we consider the Gaussian distribution with fixed mean value $\mu=0$ and varied standard deviation $\Delta$. The blue and red curves correspond to the $N=2$ and $3$ cases, respectively.}
    \label{fig_global2local}
\end{figure}

Lastly, we examine the influence of different priors $p(\theta)$ on the information $\mathcal{J}$ gained in a global estimation strategy when the input state $\rho$ is chosen as the GHZ state $\left( |0\rangle ^{\otimes N}+|1\rangle ^{\otimes N} \right) /\sqrt{2}$. Hereafter we use $\mathcal{J}(\rho)$ to represent the information gained with this particular choice of input state. We consider the Gaussian distribution with mean $\mu=0$ and varied standard deviation $\Delta$. Intuitively speaking, a smaller $\Delta$ corresponds to more prior knowledge about $\theta$. Figure \ref{fig_global2local} shows $\mathcal{J}(\rho)/\mathcal{J}_{\max}^{(i)}$ as a function of $\Delta$ for global parallel strategies. The blue and red curves correspond to $N=2$ and $3$, respectively. As can be seen from Fig.~\ref{fig_global2local}, $\mathcal{J}(\rho)/\mathcal{J}_{\max}^{(i)}$ is approximately $1$ for a small $\Delta$ in both the $N=2$ and $3$ cases. This is consistent with the known result that the GHZ is optimal for local parallel strategies. However, $\mathcal{J}(\rho)/\mathcal{J}_{\max}^{(i)}$ decreases as $\Delta$ increases. This means that the GHZ state is no longer optimal when there is no substantial prior knowledge about $\theta$. We also examined the impact of different priors on the performance of global sequential strategies. As the numerical results are similar to those in Fig.~\ref{fig_global2local}, we omitted them here.

\end{document}